\documentclass[journal, twocolumn]{IEEEtran}
\usepackage{color}
\usepackage{graphicx, pstool}
\usepackage{amsmath}
\usepackage{amsfonts}
\usepackage{amssymb}
\usepackage{booktabs}
\usepackage{dsfont}
\usepackage{cases}
\usepackage{tabularx}
\usepackage{boxedminipage}
\usepackage{url}
\usepackage{mathrsfs}
\usepackage{cite}
\usepackage{authblk}
\usepackage{amsmath}
\usepackage{setspace}
\usepackage{mathtools}
\usepackage{breqn}

\usepackage{algorithm}
\usepackage[noend]{algpseudocode}
\newtheorem{theorem}{Theorem}
\newtheorem{remark}{Remark}

\newtheorem{proposition}{Proposition}

\usepackage{eurosym}
\begin{document}

\title{\huge Dynamic Time-Frequency Division Duplex}
\author{Mohsen Mohammadkhani Razlighi,\textit{ Student Member, IEEE}, Nikola Zlatanov, \textit{ Member, IEEE}, and Petar Popovski, \textit{ Fellow, IEEE}
\thanks{}
\thanks{This work has been published in part at  IEEE ICC 2018 \cite{8403595}.}
\thanks{This work was supported by the Australian Research Council Discovery Project under Grant DP180101205.}
\thanks{Content presented in this article is subjected to Australian Provisional Patent Application 2019903224.}
\thanks{M. M. Razlighi  N. Zlatanov are  with the Department of Electrical and Computer Systems Engineering, Monash University, Melbourne, VIC 3800, Australia (e-mails:  mohsen.mohammadkhanirazlighi@monash.edu and  nikola.zlatanov@monash.edu.)

P. Popovski is with the Department of Electronic Systems, Aalborg University, Denmark (e-mail: petarp@es.aau.dk).
}
}
\maketitle

\begin{abstract}
 In this paper, we introduce dynamic time-frequency-division duplex (D-TFDD), which is a novel duplexing scheme that combines time-division duplex (TDD) and frequency-division duplex (FDD). In D-TFDD, a user receives from the base station (BS) on the downlink in one frequency band and transmits to the BS on the uplink in another frequency band, as in FDD. Next, the user shares its uplink  transmission (downlink reception) on the corresponding frequency band with the uplink  transmission  or the downlink  reception   of another user in a D-TDD fashion. Hence, in a given frequency band, the BS communicates with user 1 (U1)  and  user 2 (U2) in a D-TDD fashion. The proposed D-TFDD scheme does not require inter-cell interference (ICI) knowledge  and only requires   channel state information (CSI) of the local BS-U1 and BS-U2 channels. Thereby, it is practical for implementation. The proposed D-TFDD scheme increases the throughput region between the BS and the two users in a given frequency band, and significantly decreases the outage probabilities on the corresponding BS-U1 and BS-U2 channels. Most importantly, the proposed D-TFDD scheme doubles  the diversity gain on both the corresponding BS-U1 and the BS-U2 channels  compared to the diversity gain of existing duplexing schemes, which results in very large performance gains.
\end{abstract}

\section{Introduction}\label{Sec-Intro}
\begin{figure}
\centering
 \resizebox{1.1\linewidth}{!}{
\pstool[width=0.9\linewidth]{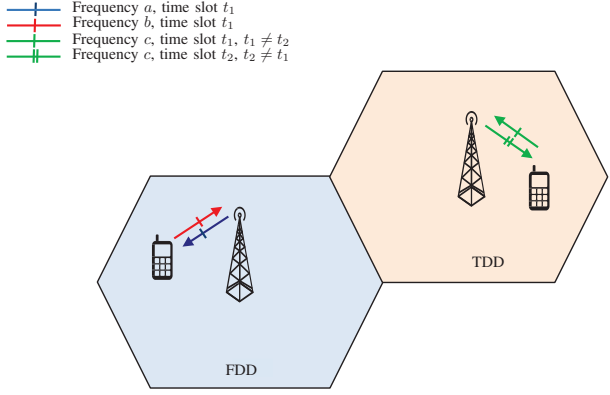}{
\psfrag{a}[ll][c][0.5]{Frequency $a$, time slot $t_1$}
\psfrag{b}[ll][c][0.5]{Frequency $b$, time slot $t_1$}
\psfrag{c}[ll][c][0.5]{Frequency $c$, time slot $t_1$, $t_1\neq t_2$}
\psfrag{d}[ll][c][0.5]{Frequency $c$,  time slot $t_2$, $t_2\neq t_1$}
\psfrag{e}[c][c][0.5]{FDD}
\psfrag{f}[ll][c][0.5]{TDD}
}}
\caption{System model of networks with FDD and TDD Communication.}
\label{Fig:sys-mod-TDD-FDD}
 \vspace*{-5mm}
\end{figure}

 Traditionally, a half-duplex (HD) base station (BS) operates in either the time-division duplex (TDD) mode or the frequency-division duplex (FDD) mode in order to receive and transmit information from/to its users. In the TDD mode, a user uses the same frequency band for uplink and downlink, while uplink and downlink transmissions occur in different time slots \cite{holma2011lte}, see Fig.~\ref{Fig:sys-mod-TDD-FDD}. On the other hand, in the FDD mode, a user is allocated two frequency bands, one dedicated for uplink and the other one for downlink, where the uplink and downlink transmissions occur simultaneously \cite{holma2011lte}, see Fig.~\ref{Fig:sys-mod-TDD-FDD}. In this paper, we  introduce  time-frequency-division duplex (TFDD), which is a novel duplexing scheme that combines TDD and FDD, and yields significant performance gains compared to TDD and FDD.

 \subsection{Background on the Different Types of Duplexing Schemes}
 \subsubsection{Static vs. Dynamic Duplexing}
  In general, the  duplexing method between a BS and its users can be  static or dynamic. In static duplexing, the time-frequency resources in which the BS performs uplink receptions and downlink transmissions from/to the user are  prefixed and unchangeable over time\cite{TechNoteLET}. On the other hand, in dynamic duplexing schemes, each time-frequency resource unit can be dynamically allocated for communications  based on the instantaneous channel state information (CSI). As a result, dynamic duplexing schemes achieve a much better performance compared to  static  duplexing schemes  \cite{7511412,6353682}, and thereby  have attracted significant research interest \cite{7511412,6353682,1386526,7876862,8004461}.

\subsubsection{Centralized Dynamic Duplexing vs. Distributed Dynamic Duplexing}
A dynamic duplexing scheme can be implemented in either centralized or distributed fashion \cite{1386526}. In centralized dynamic duplexing schemes, the decision for allocating the time-frequency resources for communication  is performed at a central node, which then informs all BSs about the decision. In this way, the communication between neighbouring cells can be synchronized in order to minimize inter-cell interference\footnote{ICI  emerges when  BSs and users in neighbouring cells transmit and receive on the same frequency band.} (ICI) 
 \cite{957300,7208806,7000558, 1336656,4769393,7876862, 7636855, 8016428, 7491359,8004461,7003998}. However, centralized dynamic duplexing schemes require at the central node full CSI   from all links in all cells in order for the central node to make an optimal  decision for allocating the time-frequency resources for each BS. In addition, the central node also needs to inform  all other network nodes about the scheduling decisions. This requires a large amount of signalling information to be exchanged between the central node and all other network nodes. As a result, implementation of centralized dynamic duplexing schemes, in most cases, is infeasible in practice.
 
 On the other hand, in distributed dynamic duplexing schemes, each BS allocates the time-frequency resources for its users without any synchronization with other BSs \cite{4769393,6666413,7070655}. To this end, only local CSI is needed at each BS. As a result, distributed dynamic duplexing schemes  are much more appropriate for practical implementation compared to the centralized dynamic duplexing scheme. However, distributed dynamic duplexing schemes have to cope with higher ICI than centralized dynamic duplexing schemes. 
 
 The proposed TFDD scheme can be characterized as a distributed dynamic duplexing scheme, which means it is suitable for  practical  implementation.

 \subsection{Contribution}
In this paper, we introduce dynamic time-frequency-division duplex (D-TFDD), which is a novel duplexing scheme that combines D-TDD  and  FDD. In D-TFDD, a user receives from the  BS  on the downlink in one frequency band and transmits to the BS on the uplink in another frequency band, as in FDD. Next, the user shares its uplink   transmission (downlink reception) on the corresponding frequency band with the uplink  transmission  or the  downlink  reception  of another user in a D-TDD fashion. Hence, in a given frequency band, the BS communicates with user 1 (U1)  and  user 2 (U2) in a D-TDD fashion. The proposed D-TFDD scheme does not require  ICI  knowledge  and only requires    CSI  of the local BS-U1 and BS-U2 channels. Thereby, it is practical for implementation. The proposed D-TFDD scheme increases the throughput region between the BS and the two users in a given frequency band, and significantly decreases the outage probabilities on the corresponding BS-U1 and BS-U2 channels. Most importantly, the proposed D-TFDD scheme doubles  the diversity gain on both the corresponding BS-U1 and the BS-U2 channels  compared to the diversity gain of existing duplexing schemes, which results in very large performance gains.

\subsection{Relevance of D-TFDD to 5G and Beyond}
One of the prominent aspects of fifth generation (5G) mobile networks is   having a flexible physical layer design. In one hand, this capability  facilitates implementing challenging physical layer protocols, and on the other hand opens the door for unconventional schemes to be implemented on the physical layer. Such a flexible hardware-software design can easily accommodate our D-TFDD scheme and thereby improve the performance of 5G networks\cite{wong_schober_ng_wang_2017}. In addition, distributed resource allocation for dense heterogeneous wireless networks is in one of the main scopes of 5G \cite{wong_schober_ng_wang_2017}, which also fits well with our D-TFDD scheme.
Moreover, proposed scheme is applicable to multi-tier multi-cell systems, which is another feature of 5G networks.

\subsection{Related Works on D-TDD and D-FDD} 

Distributed D-TDD schemes for BSs  have been investigated in \cite{6666413,7070655,1705939,4556648,1638665,7136469,4524858} and references therein. In particular, \cite{6666413} proposed cooperation among cells that resembles a centralized  D-TDD scheme. The authors in \cite{7070655} proposed a  D-TDD scheme which alleviates the ICI impairment by splitting the uplink and downlink frequency.  Authors in \cite{1705939} and \cite{4556648} investigate a distributed  D-TDD scheme designed for multiple-antennas. The work in \cite{1638665}  proposes a  distributed multi-user D-TDD scheduling scheme, where the ICI is not taken into account, which may lead to poor performance in practice. The authors in \cite{7136469} investigated an identical network  as in \cite{1638665}, but with ICI taken into account. However, the solution in \cite{7136469} is based on a brute-force search algorithm for allocating the time slots. Authors in \cite{4524858} proposed a  D-TDD scheme that  performs optimal power, rate, and user allocation. However, the ICI level in \cite{4524858} is  assumed to be fixed during all time slots, which may not be an accurate model of ICI in practice, since due to the fading and the power-allocations at neighbouring BSs, the ICI varies with time.

On the other hand, D-FDD  has been introduced in \cite{4769393}, where the authors proposed a scheme for adapting  the  downlink to uplink bandwidth ratio.

We note that \cite{4769393,6666413,7070655,1705939,4556648,1638665,7136469,4524858} require full knowledge of the ICI, which may not be practical, as discussed in Sec. \ref{Estimation_discuss}. We also note that the schemes in \cite{4769393,6666413,7070655,1705939,4556648,1638665,7136469,4524858}  transform to the static-TDD and/or static-FDD scheme when ICI is not known.  Hence,   the  static-TDD and/or static-FDD are much more practical for implementation than the  D-FDD  duplexing scheme  since they do not require ICI  knowledge.

The rest of this paper is organized as follows. In Section \ref{Sec-Sys}, we present the system and channel model. In  Sections \ref{Sec-FR} and \ref{subSec-PRL}, we present the D-TFDD schemes for the cases when the  ICI is known and unknown, respectively. Simulation and numerical results are provided in Section \ref{Sec-Num}, and the conclusions are drawn in Section \ref{Sec-Conc}.

\section{System and Channel Model}\label{Sec-Sys}
In the following, we consider a cellular network consisting of  cells, where each cell has a single BS and users that the BS serves.

\subsection{Frequency and Time Allocation in D-TFDD}

\begin{figure}
\centering
 \resizebox{1\linewidth}{!}{
\pstool[width=.9\linewidth]{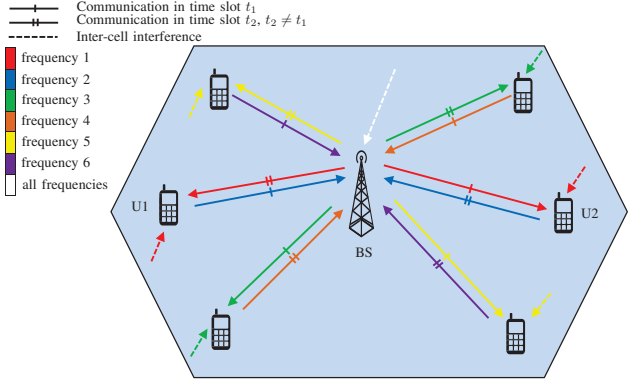}{
\psfrag{a}[ll][c][0.5]{Communication in time slot $t_1$}
\psfrag{b}[ll][c][0.5]{Communication in time slot $t_2$, $t_2\neq t_1$}
\psfrag{c}[ll][c][0.5]{Inter-cell interference}
\psfrag{g}[ll][c][0.5]{BS}
\psfrag{i}[ll][c][0.5]{U1}
\psfrag{j}[ll][c][0.5]{U2}
\psfrag{d}[c][c][0.5]{frequency 1}
\psfrag{e}[c][c][0.5]{frequency 2}
\psfrag{f}[c][c][0.5]{frequency 3}
\psfrag{h}[c][c][0.5]{frequency 4}
\psfrag{k}[c][c][0.5]{frequency 5}
\psfrag{l}[c][c][0.5]{frequency 6}
\psfrag{m}[c][c][0.5]{    \hspace{2mm}      all frequencies}
}}
\caption{Frequency Allocation Scheme 1.}
\label{Fig:sys-mod-Type23}
 \vspace*{-5mm}
\end{figure}

\begin{figure}
\centering
 \resizebox{1\linewidth}{!}{
\pstool[width=0.9\linewidth]{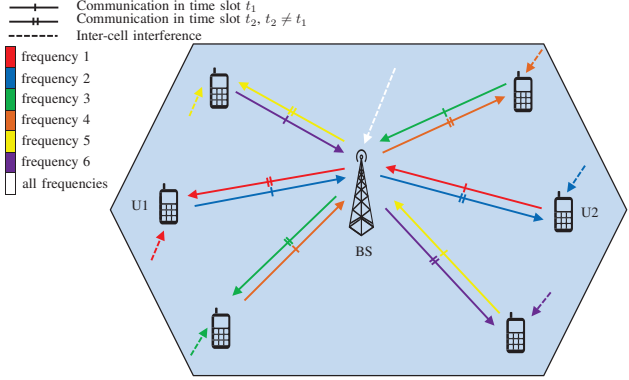}{
\psfrag{a}[ll][c][0.5]{Communication in time slot $t_1$}
\psfrag{b}[ll][c][0.5]{Communication in time slot $t_2$, $t_2\neq t_1$}
\psfrag{c}[ll][c][0.5]{Inter-cell interference}
\psfrag{g}[ll][c][0.5]{BS}
\psfrag{i}[ll][c][0.5]{U1}
\psfrag{j}[ll][c][0.5]{U2}
\psfrag{d}[c][c][0.5]{frequency 1}
\psfrag{e}[c][c][0.5]{frequency 2}
\psfrag{f}[c][c][0.5]{frequency 3}
\psfrag{h}[c][c][0.5]{frequency 4}
\psfrag{k}[c][c][0.5]{frequency 5}
\psfrag{l}[c][c][0.5]{frequency 6}
\psfrag{m}[c][c][0.5]{    \hspace{2mm}      all frequencies}
}}
\caption{Frequency Allocation Scheme 2.}
\label{Fig:sys-mod-Type1}
 \vspace*{-5mm}
\end{figure}

 In the proposed D-TFDD scheme, we have two possible frequency allocation schemes at each BS, Frequency Allocation Scheme 1 shown in Fig.~\ref{Fig:sys-mod-Type23} and Frequency Allocation Scheme 2 shown in Fig.~\ref{Fig:sys-mod-Type1}.  In both frequency allocation schemes, each user  is allocated  two distinct frequency bands whithin the cell, one for uplink transmission and the other for downlink reception,  identical as in   FDD, see Figs.~\ref{Fig:sys-mod-Type23} and \ref{Fig:sys-mod-Type1}. In Frequency Allocation Scheme~1, the  frequency band of a user allocated for uplink transmission (downlink reception) is shared in a D-TDD fashion  with the  uplink transmission (downlink reception) of another user, as shown in Figs.~\ref{Fig:sys-mod-Type23}. Whereas, in Frequency Allocation Scheme~2, the  frequency band of a user allocated for uplink transmission (downlink reception) is shared in a D-TDD fashion  with the  downlink reception (uplink transmission) of another user, as shown in Figs.~\ref{Fig:sys-mod-Type1}. Hence, for  $N$ users, the BS needs to allocate $N$ frequency bands in D-TFDD, same as in TDD.

 Frequency Allocation Scheme 1 is more appropriate for cellular communication networks, where the transmit powers of BSs are much higher than the transmit powers of the users. Specifically, when  Frequency Allocation Scheme~1 is applied to every BS in the cellular network such that the uplink frequency bands are used only for uplink and the downlink frequency bands are used only for downlink at all BSs, then all uplink links receive inter-cell interference only from other uplink transmissions and all downlink links receive inter-cell interference only from other downlink transmissions. As a result, the problem in existing D-TDD schemes of strong downlink  transmission from one BS interfering with the weak uplink reception at another BS is avoided in D-TFDD with Frequency Allocation Scheme~1. On the other hand,  Frequency Allocation Scheme 2 might be more   suitable for networks where the uplink and downlink transmissions have equal powers.  
 

\subsection{The Three-Node Subnetwork}
\begin{figure}
\centering
 \resizebox{1.1\linewidth}{!}{
\pstool[width=0.9\linewidth]{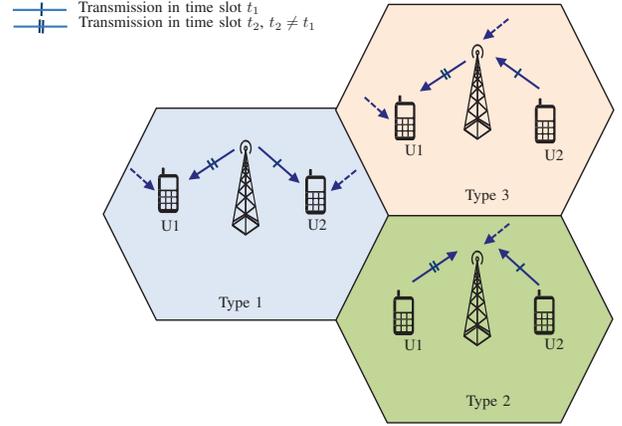}{
\psfrag{a}[ll][c][0.5]{Transmission in time slot $t_1$}
\psfrag{b}[ll][c][0.5]{Transmission in time slot $t_2$, $t_2\neq t_1$}
\psfrag{cc}[ll][c][0.5]{Inter-Cell Interference}
\psfrag{c}[ll][c][0.5]{U1}
\psfrag{e}[c][c][0.5]{Type 1}
\psfrag{f}[ll][c][0.5]{Type 3}
\psfrag{g}[ll][c][0.5]{Type 2}
\psfrag{d}[ll][c][0.5]{U2}
\psfrag{h}[ll][c][0.5]{U1}
\psfrag{l}[ll][c][0.5]{U2}
\psfrag{j}[ll][c][0.5]{U2}
\psfrag{k}[ll][c][0.5]{U1}
}}
\caption{System model of a BS and two users employing D-TFDD on a specific frequency band for Type 1, 2 and 3.}
\label{Fig:sys-mod-final}
 \vspace*{-5mm}
\end{figure}

We assume that there is no interference between different frequency bands. The only interference at a user/BS  in a given frequency band is a result of the transmission  in the same frequency band from the users and the BSs in other cells. As a result,  for a given frequency band, the considered cellular network employing the D-TFDD scheme can be divided into three-node subnetworks, where each subnetwork consists of  a BS and two users working in the same frequency band that are  impaired by ICI coming from the rest of the subnetworks working in the same frequency band, as shown in Fig.~\ref{Fig:sys-mod-final}. Depending on whether the downlink reception (uplink transmission) in a given frequency band is shared with the  downlink reception  or the  uplink transmission of another user, there can be three types of three-node subnetworks, as shown in Fig.~\ref{Fig:sys-mod-final}.  Type 1 is when both  users perform  downlink receptions in the given frequency band. Type 2 is when both  users perform  uplink transmissions in the given frequency band.  And Type 3 is when one of the users performs uplink transmission and the other user  performs downlink reception in the given frequency band.
 Note that the three types of three-node subnetworks differ only in the direction of the transmission on the BS-U1 and BS-U2 channels.

Now, in order for the D-TFDD to be a distributed duplexing scheme, each of the three-node subnetworks must perform D-TFDD  independently from the rest of the subnetworks in the cellular network. As a result, without loss of generality, the equivalent system model that needs to be investigated is comprised of a BS communicating with U1 and U2 in different time slots but in the same frequency band, where receivers are impaired by ICI, as shown in Fig.~\ref{Fig:sys-mod-final}.

\remark{
The considered three-node subnetworks  shown in Fig.~\ref{Fig:sys-mod-final}  can also represent the decoupled access \cite{7801002,7432156} by BS and users switching places, where a user is connected to two BSs and performs uplink-transmission or downlink-reception to/from BS1 and uplink-transmission or downlink-reception to/from BS2 in the same frequency band. Since the decoupled system can be obtained by the proposed system where BS and users switch places, the proposed D-TFDD scheme is also applicable to the decoupled access.
}


\subsection{Inter-Cell Interference}
The receiving nodes of a given three-node subnetwork are impaired by interference from all other nodes in the network that transmit on the  frequency band used for reception at the BS, and/or U1, and/or U2, also referred to as ICI, see Fig.~\ref{Fig:sys-mod-final}. Let the power of the ICI  at the receiving nodes on the BS-U1 and BS-U2 channels in time slot\footnote{Time slot is  a time interval that is equal or smaller than the duration of the coherence interval. Moreover, we assume that a time slot is long enough such that a capacity achieving codeword can be transmitted during one time slot.} $t$ be denoted\footnote{The subscripts 1 and 2 are used to symbolize the BS-U1   and the BS-U2 channels, respectively.} by $\gamma_{I1}(t)$ and $\gamma_{I2}(t)$, respectively. Then, we can obtain $\gamma_{I1}(t)$ and $\gamma_{I2}(t)$ as\footnote{For   D-TFDD Type 3, $\gamma_{I1}(t)=\gamma_{I2}(t)$.}
\begin{align}
\gamma_{I1}(t)&= \sum_{k \in \mathcal{K}} P_k \gamma_{k1}(t)   ,\label{eq_Ic1}\\
\gamma_{I2}(t)&= \sum_{k \in \mathcal{K}} P_k \gamma_{k2}(t) ,\label{eq_Ic2}
\end{align}
where $\mathcal{K}$  is the set of interfering nodes, $P_k$ is the power of  interfering node $k$, and   $\gamma_{k1}(t)$ and $\gamma_{k2}(t)$ are the square of the channel gains between interfering node $k$ and  the receiver on the BS-U1 channel, and interfering node $k$ and the receiver one the BS-U2 channel, in time slot $t$, respectively.

\subsection{Inter-Cell Interference Estimation Overhead}\label{Estimation_discuss}
A network comprised of  $K$ HD nodes, all operating in the same frequency band, requires at least $K$ estimation periods in order for the ICI to be estimated at all $K$ nodes. To see this, note that a HD node can either receive or transmit in a given frequency band. As a result, in order for a HD node to estimate the interference from the remaining $K-1$ HD nodes,  an estimation period must be dedicated for this purpose in which the considered HD node receives and the rest of the $K-1$ HD transmit. 
Since this process has to be repeated for each of the $K$ HD nodes, it follows that a network  comprised of $K$ HD nodes, all operating in the same frequency band, must dedicate $K$ time periods for ICI estimation at the $K$ HD nodes. Hence, ICI estimation at $K$ HD nodes entails an overhead of $K$ estimation periods. In addition, since the transmission schedule of the different HD nodes is not known in advance,  the estimated ICI may differ significantly than the real one, which  means that the overhead of $K$ time periods is a lower bound of the actual number of time periods needed for estimation of the actual ICI.  In fact, this is a key point. The only realistic way to have the transmission schedule of the  HD nodes   known in advance is to have a central controller that gathers all the channels, makes a scheduling decision for each link in each time slot and forwards that decision to the nodes. This is not feasible in current systems and will likely not be feasible in future systems as long as the coherence time equals  a time slot during which the CSI acquisition, the  transmission of the scheduling decisions, and the actual  transmission of data need to  take place. 

The overhead needed for ICI estimation requires resources that may prohibit ICI estimation   in practice.  As a result, in this paper, we investigate the practical case without ICI knowledge, and propose a distributed D-TFDD scheme for this practical scenario. In addition,  in order to obtain an  upper bound on the performance of the D-TFDD in terms of outage probability and throughput rate for unknown ICI, we will also investigate the case where the ICI is known at the nodes. Consequently, we will propose  distributed D-TFDD schemes for the cases with and without ICI knowledge, and show that the proposed distributed D-TFDD scheme without ICI knowledge has performance which is close to its upper bound achieved  when the ICI is known.

\remark{
The above method for estimating the interference works only if $K$ is known. In practice, $K$ is unknown and, in that case, estimating the interference requires even more resources.
}

\subsection{Channel Model}
In a given subnetwork, we assume that the BS-U1 and BS-U2 channels are complex-valued additive white Gaussian noise (AWGN) channels impaired by slow fading and ICI.  Next, we assume that the transmission time is divided into $ T\to\infty $ time slots. Furthermore, we assume  that the fading is constant during one time slot and changes from one time slot to the next. In time slot $t$, let the complex-valued fading gains of the BS-U1 and BS-U2 channels be denoted by   $h_{1}(t)$ and $h_{2}(t)$, respectively. Moreover, let the variances of the complex-valued AWGNs at  receiving nodes  of the the BS-U1 and BS-U2 channels be denoted by $\sigma_1^2$ and $\sigma_2^2$, respectively\footnote{For   D-TFDD Type 2, $\sigma_1^2=\sigma_2^2$.}. For convenience, we define normalized magnitude-squared fading gains of the BS-U1 and BS-U2   channels as $\gamma_{1}(t)=|h_{1}(t)|^2/\sigma_1^2$ and $\gamma_{2}(t)=|h_{2}(t)|^2/\sigma_2^2$,   respectively. Furthermore, let the transmit powers of the transmit nodes on  the BS-U1 and BS-U2 channels in time slot $t$ be denoted by $P_1$ and $P_2$, $\forall t$, respectively.

Using the above notation and taking into account the AWGNs and the ICIs given by (\ref{eq_Ic1}) and (\ref{eq_Ic2}), and treating the ICI as noise, the capacities of the BS-U1 and BS-U2 channels in time slot $t$, denoted by $C_{1}(t)$ and $C_{2}(t)$, respectively, are obtained as  
\begin{align}
C_{1}(t)& \hspace{-0.75mm} = \hspace{-0.75mm} {\log _2}\hspace{-0.75mm} \left( { \hspace{-1mm}1 \hspace{-1mm}+ \hspace{-1mm}\frac{{P_1}{\gamma _{1}}(t)}{1+\gamma _{I1}(t)}} \right)
\hspace{-1.5mm},\hspace{-0.5mm}\label{eq_c1}\\
C_{2}(t)& \hspace{-0.75mm} = \hspace{-0.75mm} {\log _2}\hspace{-0.75mm} \left( { \hspace{-1mm}1 \hspace{-1mm}+ \hspace{-1mm}\frac{{P_2}{\gamma _{2}}(t)}{1+\gamma _{I2}(t)}} \right)
\hspace{-1.5mm}.\hspace{-0.5mm}\label{eq_c2}
\end{align}

\subsection{Discrete-Rate Transmission}
We assume that the  transmit nodes on  the BS-U1 or the BS-U2 channels   transmit their codewords with rates which are  selected from discrete finite sets of data rates, denoted by $\mathcal{R}_1 = \{R_1^1,R_1^2,...,R_1^M\}$ and  $\mathcal{R}_2 = \{R_2^1,R_2^2,...,R_2^L\}$, respectively,  where $M$ and $L$ denote the total number of non-zero data rates available for transmission at the transmit nodes on  the BS-U1 and BS-U2 channels, respectively. This allows us to have a transmission model used in practice which also converges to continuous transmission rates model when $M\to\infty$ and $L\to\infty$, and to the single fixed-rate transmission model when $M=1$ and $L=1$.

\section{D-TFDD For Known ICI}\label{Sec-FR}
In this section,  we assume that the ICI is known at the nodes at the start of each time slot. Although this assumption is not practical as discussed in Sec. \ref{Estimation_discuss}, it will enable us to obtain an upper bound on the practical D-TFDD without ICI  knowledge at the nodes.

\subsection{ BS-U1 and BS-U2  Throughput Region }\label{subSec-PR}
 In a given time slot, depending on whether we are communicating on  the BS-U1 and BS-U2 channels, the considered three-node subnetwork, shown in Fig.~\ref{Fig:sys-mod-final}, can be in one   of  the following three states\\
 \textit{State 0:} No transmission occurs on both BS-U1 and BS-U2  channels.\\
\textit{State 1:}  Channel BS-U1 is selected for transmission and channel   BS-U2 is inactive/silent.\\
\textit{State 2:} Channel BS-U2 is selected for transmission and channel   BS-U1 is inactive/silent. 

In State 1, the transmitting node on the BS-U1 channel   can choose to transmit with any rate in the set $\mathcal{R}_1$. Similarly,  in State 2, the transmitting node on the BS-U2 channel   can choose to transmit with any rate in the set $\mathcal{R}_2$. In order to model these states for time slot $t$,  we introduce the binary variables $q_1^m(t)$, $m=1,2...,M$ and $q_2^l(t)$, for $l=1,...,L$, defined as
\begin{align}
q_1^m(t) =& \left\{ 
\begin{array}{ll}
1& \textrm{if channel BS-U1 is selected for the } \\
&\textrm{ transmission of a codeword with rate $R_1^m$}\\
&\textrm{ and power $P_1$  in time slot $t$}\\
0& \textrm{otherwise},
\end{array}\label{eq_q1}
 \right.\\
q_2^l(t) =& \left\{ 
\begin{array}{ll}
1& \textrm{if channel BS-U2 is selected for the } \\
&\textrm{ transmission of a codeword with rate $R_2^m$}\\
&\textrm{ and power $P_2$  in time slot $t$}\\
0& \textrm{otherwise},
\end{array}
 \right.\label{eq_q2}
\end{align} 
respectively. In addition, since the considered network can be in one and only one state in time slot $t$, the following has to hold 
\begin{align}\label{eq_PR_0}
\sum_{m=1}^M q_1^m(t)+\sum_{l=1}^L q_2^l(t)\in\{0,1\},
\end{align}
where if $\sum_{m=1}^M q_1^m(t)+\sum_{l=1}^L q_2^l(t)=0$ holds, then  both the BS-U1 and the BS-U2 channels are inactive in time slot $t$.  Condition (\ref{eq_PR_0}) results from the HD constraint of the BS, i.e., the BS can either receive  or transmit in a given time slot on the same frequency band.

Since the available transmission rates are discrete, outages can occur. An outage is defined as the event when data rate of the transmitted codeword is larger than the capacity of the underlying channel. To model the outages on the BS-U1 and the BS-U2 channels, we introduce the following auxiliary binary variables, $O_1^m(t)$, for $m=1,...,M$, and $O_2^l(t)$, for $l=1,...,L$, respectively, defined as
\begin{align}
O_1^m(t)&=\left\{
\begin{array}{ll}
1 & \textrm{if }  C_{1}(t)\geq R_1^m\\
0 & \textrm{if }  C_{1}(t)< R_1^m,
\end{array}
\right. \label{eq_PR_fr1}\\
O_2^l(t)&=\left\{
\begin{array}{ll}
1 & \textrm{if }  C_{2}(t)\geq R_2^l\\
0 & \textrm{if }  C_{2}(t)< R_2^l.
\end{array}
\right. \label{eq_PR_fr2}
\end{align}

Using $O_1^m(t)$, we can obtain that in time slot $t$  a codeword transmitted on the BS-U1 channel   with rate $R_1^m$ can be decoded correctly at the receiver if and only if (iff) $q_1^m(t)O_1^m(t)>0$ holds. Similarly,  using $O_2^l(t)$, we can obtain that in time slot $t$ a codeword ctransmitted on the BS-U2 channel    with rate $R_2^l$  can be decoded correctly at receiver iff $q_2^l(t) O_2^l(t)>0$ holds. Thereby,  the average achieved throughputs during $T\to\infty$ time slots on the BS-U1  and   BS-U2 channels, denoted  by $\bar R_{1}$ and $\bar R_{2} $, respectively,  are given by
\begin{align}
\bar R_{1} &=\lim_{T\to\infty}\frac 1T \sum_{t=1}^T \sum_{m=1}^M R_1^m   q_1^m(t) O_1^m  (t),\label{eq_PR_fr3}
\\
\bar R_{2} &=\lim_{T\to\infty}\frac 1T \sum_{t=1}^T \sum_{l=1}^L R_2^l   q_2^l(t) O_2^l  (t).\label{eq_PR_fr4}
\end{align}

The throughput pair $(\bar R_1, \bar R_2)$, defined by  (\ref{eq_PR_fr3}) and (\ref{eq_PR_fr4}), for some fixed vectors $[q_1^m(1), q_1^m(2),...,q_1^m(T)]$ and $[q_2^l(1), q_2^l(2),...,q_2^l(T)]$ gives one point on the graph where $\bar R_1$ and $\bar R_2$ are axis. All possible combinations of $[q_1^m(1), q_1^m(2),...,q_1^m(T)]$ and $[q_2^l(1), q_2^l(2),...,q_2^l(T)]$  give a region of points that is bounded by a maximum boundary line of the BS-U1  and   BS-U2   throughput region. Our task now is to find the maximum boundary line of this  BS-U1  and   BS-U2 throughput region, $(\bar R_{1} ,\bar R_{2})$, by selecting the optimal values of $q_1^m(t)$, $q_2^l(t)$,   $\forall m,l,t$, respectively. 

The maximum boundary line of the  BS-U1  and   BS-U2 throughput region $(\bar R_{1},\bar R_{2})$, given by (\ref{eq_PR_fr3}) and (\ref{eq_PR_fr4}), can be found from the following maximization problem 
\begin{align}\label{eq_op_PR}
& {\underset{q_1^m(t), q_2^l(t), \;\forall l,m,t.\;}{\textrm{Maximize: }}}\; \mu \bar R_{1}+\big(1 - {\mu} \big) \bar R_{2} \nonumber\\
 &  {\rm{Subject\;\;  to \; :}} \nonumber\\
 & \qquad {\rm C1:} \;  \ q_1^m(t)\in\{0,1\}, \forall m  \nonumber\\
 & \qquad {\rm C2:} \;  \ q_2^l(t)\in\{0,1\}, \forall l \nonumber\\
& \qquad {\rm C3:}\; \sum_{m=1}^M q_1^m(t)+\sum_{l=1}^L q_2^l(t)\in\{0,1\},
\end{align}
where $\mu$ is a priori given constant which satisfies $0\leq \mu\leq 1$. A specific value of $\mu$ provides one  point on the boundary line of  the  BS-U1  and   BS-U2 throughput region\footnote{Note that the defined throughput region is not the capacity region.} $(\bar R_{1},\bar R_{2})$. By varying $\mu$ from zero to one, the entire boundary line of the BS-U1  and   BS-U2 throughput region $(\bar R_{1},\bar R_{2})$ can be obtained. The solution of (\ref{eq_op_PR}) is given in the following theorem.

\begin{theorem}\label{Theo_PR}
The optimal state and rate selection variables, $q_1^m(t)$ and $q_2^l(t)$, of the D-TFDD scheme for known ICI that  maximize the  BS-U1  and   BS-U2  throughput region of the considered subnetwork, which are found as the solution of (\ref{eq_op_PR}), are given as
\begin{align}
&\textrm{BS-U1 transmission }  \blacktriangleright  \;\ q_1^{m^*}(t)=1,  q_1^{m}(t)=0, \; \forall m \ne m^*    \nonumber\\ 
& \qquad \qquad  \textrm{ and } q_2^l(t)=0, \; \forall l\nonumber\\ 
& \qquad \qquad \textrm{  if}\; \left[             \Lambda_1^{m^*}(t)      \geq      \Lambda_2^{l^*}(t)   \; \textrm{and}\; \Lambda_1^{m^*}(t)> 0  \right], \nonumber\\
&\textrm{BS-U2 transmission  } \blacktriangleright\; q_2^{l^*}(t)=1,  q_2^{l}(t)=0, \; \forall l \ne l^* \;    \nonumber\\
& \qquad \qquad \textrm{and}\;   q_1^m(t)=0, \; \forall m \nonumber\\ 
& \qquad \qquad\textrm{  if}\; \left[              \Lambda_2^{l^*}(t)      >      \Lambda_1^{m^*}(t)   \; \textrm{and}\; \Lambda_2^{l^*}(t) > 0  \right], \nonumber\\
&\textrm{Silence  }  \blacktriangleright  \; q_1^{m}(t)=0, \; \forall m \;\textrm{and} \; q_2^l(t)=0,   \; \forall l   \nonumber\\ 
& \qquad \qquad\textrm{  if}\; \left[             \Lambda_1^{m^*}(t)       =0     \; \textrm{and}\; \Lambda_2^{l^*}(t) = 0  \right], \label{eq_scheme_AP_G}
\end{align}
where $\Lambda_1^{m}(t)$, $\Lambda_2^{l}(t)$, $m^*$ and $l^*$ are defined as
\begin{align}
\Lambda_1^{m}(t) &=\mu R_1^{m} O_1^{m} (t), \label{eq_PR_1}\\
\Lambda_2^{l}(t) &=(1-\mu) R_2^{l} O_2^{l} (t) \label{eq_PR_2},\\
m^*&=\arg {\underset{m }{\textrm{max} }}\{\Lambda_1^{m}(t) \},\\
l^*&=\arg{\underset{l }{\textrm{max} }}\{ \Lambda_2^{l}(t) \}.
\end{align} 
\end{theorem} 
\begin{IEEEproof}
Please refer to Appendix~\ref{app_PR} for the proof.
\end{IEEEproof}

Note that for the  proposed D-TFDD scheme in Theorem~\ref{Theo_PR} to operate, the receivers of the BS-U1 and BS-U2 channels need to know  $O_1^m (t)$ and $O_2^l (t)$, respectively, $\forall m,l$, at the start of time slot $t$, which requires knowledge of the ICI.

\subsection{Diversity Gain of the Proposed D-TFDD for Known ICI}\label{sec_op}
It is quite interesting to investigate the diversity gain achieved with the  D-TFDD scheme for known ICI proposed in Theorem~\ref{Theo_PR}. In the literature, the asymptotic outage probability, from which the diversity gain is obtained, is derived assuming only a single available transmission rate at the transmitter, see \cite{1362898}. Following this convention, in the following, we  derive the asymptotic outage probabilities of the BS-U1  and the BS-U2 channels, denoted by $P_{\rm out}$, achieved with the  D-TFDD scheme for known ICI proposed in Theorem~\ref{Theo_PR} for $\mu=\frac{1}{2}$, $M=L=1$, $ P_1= P_2$, and $R_1^1=R_2^1=R_0$.  For simplicity, we only investigate the case of Rayleigh fading, and also assume that the BS-U1 and BS-U2 channels are affected by independent and identically distributed (i.i.d) fading.

\begin{theorem}\label{Theo_PR_Outage_G}
The asymptotic outage probability of the  D-TFDD scheme for known ICI proposed in Theorem~\ref{Theo_PR} for the case of Rayleigh fading and when $\mu=\frac{1}{2}$, $M=L=1$, $ P_1= P_2=P$, and $R_1^1=R_2^1=R_0$ hold, is given by
\begin{align} \label{Out_U6GG}
 P_{\rm out}& \to \frac{ \gamma_{\rm th}^2 \hat \Omega_I }{\Omega_0^2},  \; \textrm{as} \;{P\to\infty},
\end{align} 
where $\gamma_{\rm th}=\frac{2^{R_0}-1}{P} $, $\Omega_0=E\left \{\frac{|h_{1}(t)|^2}{\sigma_1^2 }\right \}=E\left\{\frac{|h_{2}(t)|^2}{\sigma_2^2}\right\}$, and $\hat \Omega_I=E \Big \{  (1+\gamma _{I1}(t)) (1+\gamma _{I2}(t)) \Big \}$.

As can be seen from (\ref{Out_U6GG}), the outage probability $ P_{\rm out}$ has a diversity gain of two.
\end{theorem} 
\begin{IEEEproof}
Please refer to Appendix~\ref{app_PR_Outage_G} for the proof.
\end{IEEEproof}

Note that  existing D-TDD and D-FDD schemes achieve a diversity gain of one, which leads to the conclusion that the proposed D-TFDD scheme doubles the diversity gain compared to existing duplexing schemes, which in turn leads to very large performance gains, cf. Sec. \ref{Sec-Num}.

\section{D-TFDD For Unknown ICI}\label{subSec-PRL}
The D-TFDD scheme  proposed in Section \ref{Sec-FR} requires the receivers of the BS-U1 and BS-U2 channels to know $O_1^m (t)$ and $O_2^l (t)$, respectively, $\forall m,l$, at the start of time slot $t$, for $t=1, 2, ... T$, which requires ICI knowledge. However, as discussed in Sec. \ref{Estimation_discuss}, the estimation of  the ICI entails  huge cost for a cellular  network comprised of $K$ HD nodes which may not be practical. Motivated by this problem, in the following, we propose a D-TFDD scheme  where the nodes do not have  knowledge of the ICI, and as a result, the network nodes do not have to waste huge resources for estimating the ICI. We only assume that the CSI of the BS-U1 and BS-U2 channels are known at the BS, i.e., we assume local CSI knowledge at the BS. Specifically, the BS knows ${\gamma _{1}}(t)$ and ${\gamma _{2}}(t)$ at the start of time slot $t$, which can be acquired by allocating two estimation periods; one for the BS-U1 channel and the other for the BS-U2 channel, which is a huge improvement compared to the $K$ estimation periods that need to be allocated when the ICI needs to be estimated, see Sec. \ref{Estimation_discuss} .

\subsection{Proposed D-TFDD For Unknown ICI } \label{SubSec_PRL_Proposed} 
The throughput region of the BS-U1 and BS-U2 channels employing the D-TFDD scheme for unknown ICI is also given by (\ref{eq_PR_fr3}) and (\ref{eq_PR_fr4}), where $O_1^m(t)$ and $O_2^l(t)$ are defined in (\ref{eq_PR_fr1}) and (\ref{eq_PR_fr2}), respectively. The only difference now is that $q_1^m(t)$ and $q_2^l(t)$ in (\ref{eq_PR_fr3}) and (\ref{eq_PR_fr4}) are different when  the D-TFDD for unknown ICI is applied.  The optimal $q_1^m(t)$ and $q_2^l(t)$, which maximize the throughput region, defined by (\ref{eq_PR_fr3}) and (\ref{eq_PR_fr4}), of the D-TFDD scheme for unknown ICI  are given in the following.

For the case when the ICI is unknown at the nodes, first we define
\begin{align}
 & m^* \buildrel \Delta \over = \arg {\underset{m }{\textrm{max} }} \{  R_1^m  O_{1,e}^m (t)\},\label{eq_AP_S134}\\
& l^* \buildrel \Delta \over  =\arg {\underset{l }{\textrm{max} }} \{  R_2^l  O_{2,e}^l (t)\}\label{eq_AP_S135},
\end{align}
where $ O_{1,e}^m (t)$ and $O_{2,e}^l (t)$ are defined as
\begin{align}
O_{1,e}^m(t)&=\left\{
\begin{array}{ll}
1 & \textrm{if }  C_{1}^e(t)\geq R_1^m\\
0 & \textrm{if }  C_{1}^e(t)< R_1^m,
\end{array}
\right. \label{eq_PR_fr11l}\\
O_{2,e}^l(t)&=\left\{
\begin{array}{ll}
1 & \textrm{if }  C_{2}^e(t)\geq R_2^l\\
0 & \textrm{if }  C_{2}^e(t)< R_2^l.
\end{array}
\right. \label{eq_PR_fr22l}
\end{align}
The variables,  $C_{1}^e(t)$ and $C_{2}^e(t)$, used in (\ref{eq_PR_fr11l}) and (\ref{eq_PR_fr22l}),  are defined as
\begin{align}
C_{1}^e(t)&={\log _2}\left( {1 + \frac{{P_1}{\gamma _{1}}(t)}{1+\gamma_{I1}^e(t)}} \right)   
,\label{eq_c11}\\
C_{2}^e(t)&={\log _2}\left( {1 + \frac{{P_2}{\gamma _{2}}(t)}{1+\gamma_{I2}^e(t)}} \right)
,\label{eq_c22}
\end{align}
where $\gamma_{I1}^e(t)$ and $\gamma_{I2}^e(t)$ are given in Proposition~\ref{Theo_PR_L0cal} in the following, and they can be thought of as estimates of the ICI in time slot $t$.

The optimal $q_1^m(t)$ and $q_2^l(t)$, which maximize the throughput region, defined by (\ref{eq_PR_fr3}) and (\ref{eq_PR_fr4}), of the D-TFDD for unknown ICI  are as follows
\begin{align}
&\textrm{BS-U1 transmission   } \blacktriangleright \; q_1^{m^*}(t)=1,  q_1^{m}(t)=0, \; \forall m \ne m^*   \nonumber\\ 
& \qquad \qquad \textrm{ and } q_2^l(t)=0, \; \forall l,\nonumber\\ 
& \qquad \qquad\textrm{  if}\; \left[      \Lambda_1(t)       \geq     \Lambda_2(t)  \; \textrm{and}\; \Lambda_1(t) > 0  \right], \nonumber\\
&\textrm{BS-U2 transmission   } \blacktriangleright  \;q_2^{l^*}(t)=1, q_2^{l}(t)=0, ; \forall l \ne l^*,   \nonumber\\ 
& \qquad \qquad \textrm{ and } q_1^m(t)=0, \; \forall m,\nonumber\\ 
& \qquad \qquad\textrm{  if}\; \left[      \Lambda_2(t)       \geq     \Lambda_1(t)  \; \textrm{and}\; \Lambda_2(t) > 0  \right], \nonumber\\
&\textrm{Silence  } \blacktriangleright  \; q_1^{m}(t)=0, \forall m \;\textrm{and} \; q_2^l(t)=0, \;  \; \forall l, \nonumber\\ 
& \qquad \qquad\textrm{  if}\; \left[     \Lambda_1(t)      =0     \; \textrm{and}\; \Lambda_2(t) = 0  \right], \label{eq_scheme_AP}
\end{align}
where  $\Lambda_1(t)$  and $\Lambda_2(t)$ are given by
\begin{align}
\Lambda_1(t) &= \mu C_{1}^e(t),\label{eq_AP_S1111}\\
\Lambda_2(t) &= (1-\mu) C_{2}^e(t),\label{eq_AP_S2222}
\end{align}
and $m^*$ and $l^*$ are given by (\ref{eq_AP_S134}) and (\ref{eq_AP_S135}), respectively.
In (\ref{eq_AP_S1111}) and (\ref{eq_AP_S2222}), $\mu$ is a priori given constant which satisfies $0\leq \mu\leq 1$. By varying $\mu$ from zero to one, the entire boundary line of the BS-U1 and BS-U2 throughput region can be obtained.

\begin{proposition}\label{Theo_PR_L0cal}
The variables $\gamma_{I1}^e(t)$ and $\gamma_{I2}^e(t)$, found in the expressions in (\ref{eq_c11}) and (\ref{eq_c22}), which maximize the BS-U1 and BS-U2 throughput region of the D-TFDD scheme for unknown ICI proposed in (\ref{eq_scheme_AP}) are found as follow
\begin{align}
&\gamma_{I1}^e(t+1)=\gamma_{I1}^e(t) -\delta_1(t) \Phi_1(t),\label{eq_z1_Rsr1}\\
&\gamma_{I2}^e(t+1)=\gamma_{I2}^e(t) -\delta_2(t) \Phi_2(t),\label{eq_z1_Rsr2}
\end{align}
where $\delta_{k}( t)$, for $k\in\{1,2\}$,  can be some properly chosen monotonically decaying function of $t$ with $\delta_{k}( 1)<1$, such as $\frac{1}{2t}$. Furthermore,  $\Phi_1(t)$  and $\Phi_2(t)$ in (\ref{eq_z1_Rsr1}) and (\ref{eq_z1_Rsr2}) are  obtained as
\begin{align} 
&\Phi_1(t)\hspace{-0.5mm}=\hspace{-0.5mm}\frac{t\hspace{-0.5mm}-\hspace{-0.5mm}1}{t} \Phi_1(t\hspace{-0.5mm}-\hspace{-0.5mm}1) \hspace{-0.5mm}\label{eq_zeta_Rsr1131}\\
&-\hspace{-0.75mm}\frac{1}{t}  \frac{  P_1 \gamma_{1}(t) \delta_1^{m^*} \hspace{-0.75mm}(t) q_1^{m^*} \hspace{-0.75mm}(t)\bigg (  \hspace{-0.75mm} 2\chi_1^e(t) \big [ O_{1,e}^{m^*}(t)\hspace{-0.75mm}-\hspace{-0.75mm}O_{1}^{m^*}\hspace{-0.75mm}(t) \big]\hspace{-0.75mm}-\hspace{-0.75mm}\mu R_1^{m^*}    \hspace{-0.75mm} \bigg)}{\ln(2) \Big (1+\gamma_{I1}^e(t)+P_1 \gamma _{1}(t)\Big) \Big (1+\gamma_{I1}^e(t)\Big)}, \nonumber \\
&\Phi_2(t)\hspace{-0.5mm}=\hspace{-0.5mm}\frac{t\hspace{-0.5mm}-\hspace{-1.5mm}1}{t} \Phi_2(t\hspace{-1mm}-\hspace{-0.5mm}1) \hspace{-0.5mm}\label{eq_zeta_Rsr1231}\\
&-\hspace{-0.5mm}\frac{1}{t} \frac{  P_2 \gamma_{2}(t) \delta_2^{l^*}\hspace{-0.75mm}(t) q_2^{l^*}\hspace{-0.75mm}(t) \hspace{-0.75mm}\bigg (  \hspace{-1mm} 2 \chi_2^e(t) \big [ O_{2,e}^{l^*}(t)\hspace{-0.75mm}-\hspace{-.75mm} O_{2}^{l^*}\hspace{-0.75mm}(t) \big] \hspace{-0.75mm}-\hspace{-0.75mm}(1\hspace{-1.25mm}-\hspace{-1mm}\mu) R_2^{l^*}   \hspace{-1.25mm} \bigg)}{\ln(2) \Big (1+\gamma_{I2}^e(t)+P_2 \gamma _{2}(t)\Big) \Big (1+\gamma_{I2}^e(t)\Big)}, \nonumber 
\end{align}
respectively, where $\delta_1^{m^*}(t)$  and $\delta_2^{l^*}(t)$, are defined as
\begin{align}
\delta_1^{m^*} \hspace{-0.75mm}(t) \hspace{-0.75mm}=\hspace{-0.75mm}\left\{
\begin{array}{ll}
\hspace{-1.75mm}1\hspace{-1.75mm} & \hspace{-1.75mm}\textrm{if } \hspace{-0.75mm} \left(R_1^{m^*}\hspace{-0.75mm}-C_{1}^e(t\hspace{-0.75mm}-\hspace{-0.75mm}1) \right) \hspace{-0.75mm}\left(R_1^{m^*}\hspace{-0.75mm}-\hspace{-0.75mm}C_{1}^e(t)\right)\hspace{-0.75mm} \leq \hspace{-1.25mm}0 \label{eq_2elta_fr1}\\
\hspace{-1.75mm}0 \hspace{-1.75mm}& \hspace{-1.75mm}\textrm{if } \hspace{-0.75mm} \left(R_1^{m^*}\hspace{-0.75mm}-C_{1}^e(t\hspace{-0.75mm}-\hspace{-0.75mm}1) \right) 
\hspace{-0.75mm}\left(R_1^{m^*}\hspace{-0.75mm}- \hspace{-0.75mm}C_{1}^e(t)\right)\hspace{-0.75mm}>\hspace{-1.25mm}0,
\end{array}
\right.
\end{align}
and
\begin{align}\label{eq_2elta_fr2}
\delta_2^{l^*}(t)=\left\{
\begin{array}{ll}
\hspace{-1.75mm}1\hspace{-1.75mm} & \hspace{-1.75mm}\textrm{if }  \hspace{-0.75mm}\left(R_2^{l^*}\hspace{-0.75mm}-\hspace{-0.75mm}C_{2}^e(t\hspace{-0.75mm}-\hspace{-0.75mm}1) \right) \left(R_2^{l^*}\hspace{-0.75mm}-\hspace{-0.75mm}C_{2}^e(t)\right)\hspace{-0.75mm} \leq\hspace{-0.75mm} 0\\
\hspace{-1.75mm}0\hspace{-1.75mm} & \hspace{-1.75mm}\textrm{if }  \hspace{-0.75mm}\left(R_2^{l^*}\hspace{-0.75mm}-\hspace{-0.75mm}C_{2}^e(t\hspace{-0.75mm}-\hspace{-0.75mm}1) \right)
\left(R_2^{l^*}\hspace{-0.75mm}-\hspace{-0.75mm} C_{2}^e(t)\right) \hspace{-0.75mm}>\hspace{-0.75mm}0,
\end{array}
\right.
\end{align}
respectively. On the other hand, the variables  $\chi_1^e(t)$ and $\chi_2^e(t)$  in (\ref{eq_zeta_Rsr1131}) and (\ref{eq_zeta_Rsr1231}) are calculated  as
\begin{align}
&\chi_k^e(t+1)=\chi_k^e(t) +\delta_k^\chi( t) \left[ {\bar \epsilon}_k(t) - \epsilon\right]^+, \, k\in\{1,2\},\label{eq_z1_Rsr1234}
\end{align}
where $\delta_k^\chi( t)$ for $k\in\{1,2\}$ can be some properly chosen monotonically decaying function of $t$ with $\delta_k^\chi( 1)<1$, such as $\frac{1}{2t}$, $\epsilon$ is a small constant which can be configured at the system level, and $[.]^+$ denotes only positive values. Moreover, in (\ref{eq_z1_Rsr1234}), ${\bar \epsilon}_k(t)$, for $k\in\{1,2\}$,  are obtained as 
\begin{align} 
{\bar \epsilon}_1(t)&\hspace{-0.75mm}=\frac{t\hspace{-0.75mm}-\hspace{-0.75mm}1}{t} {\bar \epsilon}_U(t\hspace{-0.75mm}-\hspace{-0.75mm}1) \hspace{-0.75mm}+\hspace{-0.75mm}\frac{1}{t} q_1^{m^*}(t) \bigg( O_{1}^{m^*}(t)-O_{1,e}^{m^*}(t)  \bigg )^2\hspace{-2mm},\label{eq_zeta_Rsr1122}\\
{\bar \epsilon}_2(t)&\hspace{-0.75mm}=\frac{t\hspace{-0.75mm}-\hspace{-0.75mm}1}{t} {\bar \epsilon}_D(t\hspace{-0.75mm}-\hspace{-0.75mm}1) \hspace{-0.75mm}+\hspace{-0.75mm}\frac{1}{t} q_2^{l^*}(t) \bigg( O_{2}^{l^*}(t)-O_{2,e}^{l^*}(t)  \bigg )^2\hspace{-2mm}.\label{eq_zeta_Rsr1133}
\end{align}
Finally, we note that, ${\bar \epsilon}_k(0)$, $\Phi_k(0)$, $\chi_k^e(0)$, and $\gamma_{Ik}^e(0)$, $k \in\{1,2\}$, found in (\ref{eq_z1_Rsr1})-(\ref{eq_zeta_Rsr1133}), are initialized  to zero. The variables $\Phi_1(t)$, $\Phi_2(t)$, $\chi_1^e(t+1)$,  $\chi_2^e(t+1)$, $\delta_1^{m^*}(t)$,   $\delta_2^{l^*}(t)$, ${\bar \epsilon}_1(t)$ and ${\bar \epsilon}_2(t)$ are auxiliary variables used for real-time estimation of the ICIs $\gamma_{I1}^e(t)$ and $\gamma_{I2}^e(t)$.

The D-TFDD scheme for unknown ICI is provided in an algorithmic form in  Algorithm~\ref{Lyp_algorithm}.

\begin{algorithm}
\centering
\caption{Finding the optimal decision variables, $q_1^{m},\forall m$, and $q_2^l(t),\forall l$, for a given value of $\mu$.}
\label{Lyp_algorithm}
\begin{algorithmic}[1]
\State Initiate ${\bar \epsilon}_k(0)$, $\Phi_k(0)$, $\chi_k^e(1)$, and $\gamma_{Ik}^e(1)$, $k \in\{1,2\}$  to zero
\Procedure{ $\forall \;t \in \{1,2,...,T\}$}{}\label{lyp_proc}
\State \textrm{****** Start of time slot $t$*****}
\State compute $C_{1}^e(t)$ and $C_{2}^e(t)$ with (\ref{eq_c11}) and (\ref{eq_c22});
\State compute $O_{1,e}^m(t)$ and $O_{2,e}^l(t)$ with (\ref{eq_PR_fr11l}) and (\ref{eq_PR_fr22l});
\State compute $m^* $ and $l^* $ with (\ref{eq_AP_S134}) and (\ref{eq_AP_S135});
\State compute $\Lambda_1(t)$ and $\Lambda_2(t)$ with (\ref{eq_AP_S1111}) and (\ref{eq_AP_S2222});
\State compute $q_1^{m},\forall m$ and $q_2^l(t),\forall l$ with (\ref{eq_scheme_AP});
\State \textrm{******Scheduling Decisions Sent for Execution*****}
\State \textrm{******Transmission of data begins*****}
\State \textrm{******Transmission of data ends*****}
\State \textrm{******Feedback, $O_{1}^{m^*}(t)$/$ O_{2}^{l^*}(t)$, Received*****}
\State update ${\bar \epsilon}_1(t)$ and ${\bar \epsilon}_2(t)$ with (\ref{eq_zeta_Rsr1122}) and (\ref{eq_zeta_Rsr1133});
\State compute $\delta_1^{m^*}(t)$ and $\delta_2^{l^*}(t)$ with (\ref{eq_2elta_fr1}) and (\ref{eq_2elta_fr2});
\State update $\Phi_1(t)$ and $\Phi_2(t)$ with (\ref{eq_zeta_Rsr1131}) and (\ref{eq_zeta_Rsr1231});
\State \textrm{******Estimation for Next Time Slot, $t+1$*****}
\State update $\chi_1^e(t+1)$ and $\chi_2^e(t+1)$ with (\ref{eq_z1_Rsr1234});
\State update $\gamma_{I1}^e(t+1)$ and $\gamma_{I2}^e(t+1)$ with (\ref{eq_z1_Rsr1}) and (\ref{eq_z1_Rsr2});
\State \textrm{****** End of time slot $t$*****}
\EndProcedure
\end{algorithmic}
\end{algorithm}

\end{proposition} 
\begin{IEEEproof}
Please refer to Appendix~\ref{app_PR_Local} for the proof.
\end{IEEEproof}

\remark{
All equations from (\ref{eq_AP_S134}) to (\ref{eq_zeta_Rsr1133}) are straightforward calculations that do not depend on any hidden function or loop routines. As a result, the complexity of Algorithm 1 grows linearly with $M$ or $L$.}

\remark{
Note that for the D-TFDD scheme for unknown ICI proposed above, the functions $\Phi_1(t)$ and $\Phi_2(t)$ need to be calculated at the end of time slot $t$. To this end, note that all the variables in (\ref{eq_zeta_Rsr1131}) and  (\ref{eq_zeta_Rsr1231}) are known at the end of time time slot $t$. In particular, the  outage variables $O_{1}^{m^*}(t)$ and $O_{2}^{l^*}(t)$ either are already available at the BS if the BS is the receiver node at the end of time slot $t$, or are computed at the BS by a 1-bit  feedback from the transmitting node. Moreover,  $O_{1}^{m^*}(t)=1$  if the message received by the receiving node of the BS-U1 channel is decoded successfully  in time slot $t$, otherwise $O_{1}^{m^*}(t)=0$. Similarly,   $O_{2}^{l^*}(t)=1$  if the message received by the receiving node of the BS-U2 channel is decoded successfully  in time slot $t$, otherwise $O_{2}^{l^*}(t)=0$. Hence, the BS is able to calculate (\ref{eq_zeta_Rsr1131}) and  (\ref{eq_zeta_Rsr1231}) at the end of time slot $t$,  compute $\gamma_{I1}^e(t+1)$ and $\gamma_{I2}^e(t+1)$ by using (\ref{eq_z1_Rsr1}) and (\ref{eq_z1_Rsr2}) at the end of time slot t, and use it at the start of time slot $t+1$ by plugging in (\ref{eq_c11}) and (\ref{eq_c22}), then plug (\ref{eq_c11}) and (\ref{eq_c22}) into  (\ref{eq_AP_S1111}) and (\ref{eq_AP_S2222}), respectively, and thereby be able to compute the selection variables in (\ref{eq_scheme_AP}).}

\remark{
If the statistical characteristics of the ICI remain constant over time, $\gamma_{I1}^e(t)$ and $\gamma_{I2}^e(t)$ in (\ref{eq_z1_Rsr1}) and (\ref{eq_z1_Rsr2})  converge to  constants such as $\mathop {\lim }\limits_{t \to \infty } \gamma_{I1}^e(t) =\gamma_{I1}$ and $\mathop {\lim }\limits_{t \to \infty } \gamma_{I2}^e(t) =\gamma_{I2}$. On the other hand, if the statistical characteristics of the ICI is slowly changing over time, the proposed D-TFDD   still works.}

\subsection{Diversity Gain of the Proposed D-TFDD for Unknown ICI }\label{subSec-OUT_L}
It is quite interesting to investigate the diversity gain that the D-TFDD scheme for unknown ICI, proposed in Section~\ref{SubSec_PRL_Proposed}, achieves.  In the following, we  derive the asymptotic outage probabilities of the  BS-U1  and the BS-U2 channels, achieved with the  D-TFDD scheme for unknown ICI proposed in Sec.~\ref{SubSec_PRL_Proposed} for the case  when $\mu=\frac{1}{2}$, $M=L=1$, and $R_1^1=R_2^1=R_0$.

\begin{theorem}\label{Theo_PR_Outage_L}
The asymptotic outage probability of the  D-TFDD scheme for unknown ICI,  proposed in Section~\ref{SubSec_PRL_Proposed}, for the case of Rayleigh fading and when $\mu=\frac{1}{2}$, $M=L=1$, $ P_1= P_2=P$, and $R_1^1=R_2^1=R_0$ hold, is given by
\begin{align}\label{eq_out1asl}
P_{\rm out}& \to \frac{\gamma_{\rm th}^2}{\Omega_0^2} \left[  \hat \Omega_{I1} + \hat \Omega_{I2} + \hat \Omega_{IS} \right], \; \textrm{as} \;P \to \infty,
\end{align} 
where $\gamma_{\rm th}=\frac{2^{R_0}-1}{P} $, $\Omega_0=E\left\{\frac{|h_{1}(t)|^2}{\sigma_1^2 (1+\gamma_{I1})}\right\}=E\left\{\frac{|h_{2}(t)|^2}{\sigma_2^2 (1+\gamma_{I2})}\right\}$,
\begin{align}
&{\hat \Omega_{I1}} \hspace{-0.5mm}=\hspace{-0.5mm} \Omega_I \hspace{-0.5mm} \left(\hspace{-0.5mm}\frac{1\hspace{-0.5mm}+\hspace{-0.5mm}\gamma _{I2}}{1\hspace{-0.5mm}+\hspace{-0.5mm}\gamma _{I1}} \right)\hspace{-0.5mm} \left( \sum_{n=0}^{K-1} [1\hspace{-0.5mm}+\hspace{-0.5mm}(n\hspace{-0.5mm}+\hspace{-0.5mm}1) \Omega_I ]e^{-{\frac{\gamma _{I1}}{\Omega_I}}}   \sum_{t=0}^{n}  \frac{({\frac{\gamma _{I1}}{\Omega_I}})^i}{t!} \hspace{-0.5mm} \right), \label{Otage_U_final}\\
&{\hat \Omega_{I2}} \hspace{-0.5mm}=\hspace{-0.5mm} \Omega_I \hspace{-0.5mm} \left(\hspace{-0.5mm}\frac{1\hspace{-0.5mm}+\hspace{-0.5mm}\gamma _{I1}}{1\hspace{-0.5mm}+\hspace{-0.5mm}\gamma _{I2}} \right)\hspace{-0.5mm} \left( \sum_{n=0}^{K-1} [1\hspace{-0.5mm}+\hspace{-0.5mm}(n\hspace{-0.5mm}+\hspace{-0.5mm}1) \Omega_I ]e^{-{\frac{\gamma _{I2}}{\Omega_I}}}   \sum_{t=0}^{n}  \frac{({\frac{\gamma _{I2}}{\Omega_I}})^i}{t!} \hspace{-0.5mm} \right), \label{Otage_D_final}\\
&{\hat \Omega_{IS}} = ({1+\gamma _{I1}}) ({1+\gamma _{I2}}),\label{Otage_S_final} \\
&\Omega_I=E\{\gamma _{I1}(t)\}=E\{\gamma _{I2}(t)\}.
\end{align}
In the above expression, $\gamma_{I1}$ and $\gamma_{I2}$ are constant values obtained as $\mathop {\lim }\limits_{t \to \infty } \gamma_{I1}^e(t) =\gamma_{I1}$ and $\mathop {\lim }\limits_{t \to \infty } \gamma_{I2}^e(t) =\gamma_{I2}$. It is clear from (\ref{eq_out1asl}) that $ P_{\rm out}$ has a diversity gain of two.

\end{theorem} 
\begin{IEEEproof}
Please refer to Appendix~\ref{app_PR_Outage_L} for the proof.
\end{IEEEproof}

The result in Theorem~\ref{Theo_PR_Outage_L} shows that the  D-TFDD scheme for unknown ICI proposed in Sec. \ref{SubSec_PRL_Proposed}  achieves double the diversity gain  compared to  existing duplexing schemes, which leads to very large performance gains. Moreover, Theorem~\ref{Theo_PR_Outage_L}  shows that doubling of the diversity  gain on both the BS-U1 and BS-U2 channels is achievable even when there is no  ICI knowledge at the nodes, which is a very interesting result that shows the superior performance of the  D-TFDD scheme for unknown ICI proposed in Sec. \ref{subSec-PR} compared to existing duplexing schemes.

\section{Simulation And Numerical Results}\label{Sec-Num}
In this section, we evaluate the performance of the proposed D-TFDD scheme for unknown ICI and its upper bound, the proposed D-TFDD scheme for known ICI. Then, we compare its performance to the performance achieved with a static-TDD scheme and to the state-of-the-art D-TDD and D-FDD schemes. To this end, we first introduce the benchmark schemes and then present the numerical results.


\subsection{Benchmark Schemes}
\subsubsection{Static-TDD}
In the static-TDD scheme, see \cite{holma2011lte}, the BS  receives  and transmits in prefixed time slots.  Assuming  single transmission rates at the transmitting nodes of the BS-U1 and BS-U2 channels, and assuming that that the fractions of the total number of time slots, $T$, allocated on the BS-U1 and BS-U2 channels are $\mu$ and $1-\mu$, respectively, e.g., channel BS-U1 is active in the first $\mu T$ and channel BS-U2 is active in following $(1-\mu) T$ time slots, the BS-U1 and BS-U2 throughput during $T\to\infty$ time slots, is given by
\begin{align}
\bar R_{k}&=\lim_{T\to\infty}\frac {\mu_k}{T} \sum_{t=1}^T  O_k^1(t) R_k^1, k\in\{1,2\},\label{eq_ben_1}
\end{align}
and the outage probability is given by 
\begin{align}
P_{\rm out}&=1- \lim_{T\to\infty}\left ( \frac {1}{T} \sum_{t=1}^{\mu T}  O_1^1(t) +\sum_{t=\mu T+1}^{T} O_2^1(t) \right ),\label{eq_ben_123}
\end{align}
where $O_k^1(t)$, $k\in\{1,2\}$, is defined as
\begin{align}
O_k^1(t)&=\left\{
\begin{array}{ll}
1 & \textrm{if }   \log_2\bigg(1+\frac{ P_k}{\mu_k} \gamma_{k}(t)\bigg)\geq R_k^1\\
0 & \textrm{if }   \log_2\bigg(1+\frac{ P_k}{\mu_k} \gamma_{k}(t)\bigg)< R_k^1.
\end{array}\label{eq_ben_fr1}
\right. 
\end{align}

\subsubsection{D-TDD Scheme}
The distributed D-TDD scheme  proposed in \cite{4524858} is considered as a benchmark for comparison. We note that this scheme requires full knowledge of the ICI, which, as argued  in Section \ref{Estimation_discuss}, is not practical. In addition, we note that this scheme  without ICI knowledge transforms to the static-TDD.  Hence, the practical distributed TDD scheme is the static-TDD since it does not need ICI  knowledge. The distributed D-TDD scheme  in \cite{4524858} can serve only as an upper bound to the practical static-TDD.

 Note that D-TDD and D-FDD schemes  only different in how they share the time and frequency resources, but achieve the same performance. Therefore, for comparison purposes, we can use either of them. In the following,  we choose the D-TDD scheme.

\subsection{Numerical Results}
  All of the  results presented in this section have been performed by numerical evaluation of the derived results and are confirmed by Monte Carlo simulations. Moreover, Rayleigh fading for the BS-U1 and  BS-U2 channels, and Chi-square distribution for the ICI at the receiving nodes of the BS-U1 and  BS-U2 channels are assumed. In all the  numerical examples in Figs.~\ref{Outage_Discrete_1}-\ref{discrete_rate_SNR}, we assume   $M = L$  and $R_1^k=R_2^k=kR$, for $k = 1, 2, ..., M$, where $R$ is defined differently depending on the corresponding example.  The signal to interference plus noise ratio (SINR) is defined as the ratio of the average received signal power to interference power plus the noise power.

\subsubsection{Constraint on the Average Transmit Power} 
In the  numerical examples,  we select the fixed powers $P_1$ and $P_2$ such that the following long-term power constraints hold 
\begin{align}
&\lim_{T\to\infty}\frac 1T \sum_{t=1}^T \sum_{m=1}^M q_1^m(t)P_1 \leq \bar P_1\;\nonumber\\
\textrm{and}\; &\lim_{T\to\infty}\frac 1T \sum_{t=1}^T \sum_{l=1}^L q_2^l(t)P_2 \leq \bar P_2.\label{eq_PR_p2}
\end{align}
This enables all duplexing schemes to use identical average transmit powers, and thereby enable fair comparison between the different schemes.

\subsubsection{Outage Probability}\label{Sec-Out}
\begin{figure}[t]
\vspace*{-2mm}
\centering\includegraphics[width=3.4in]{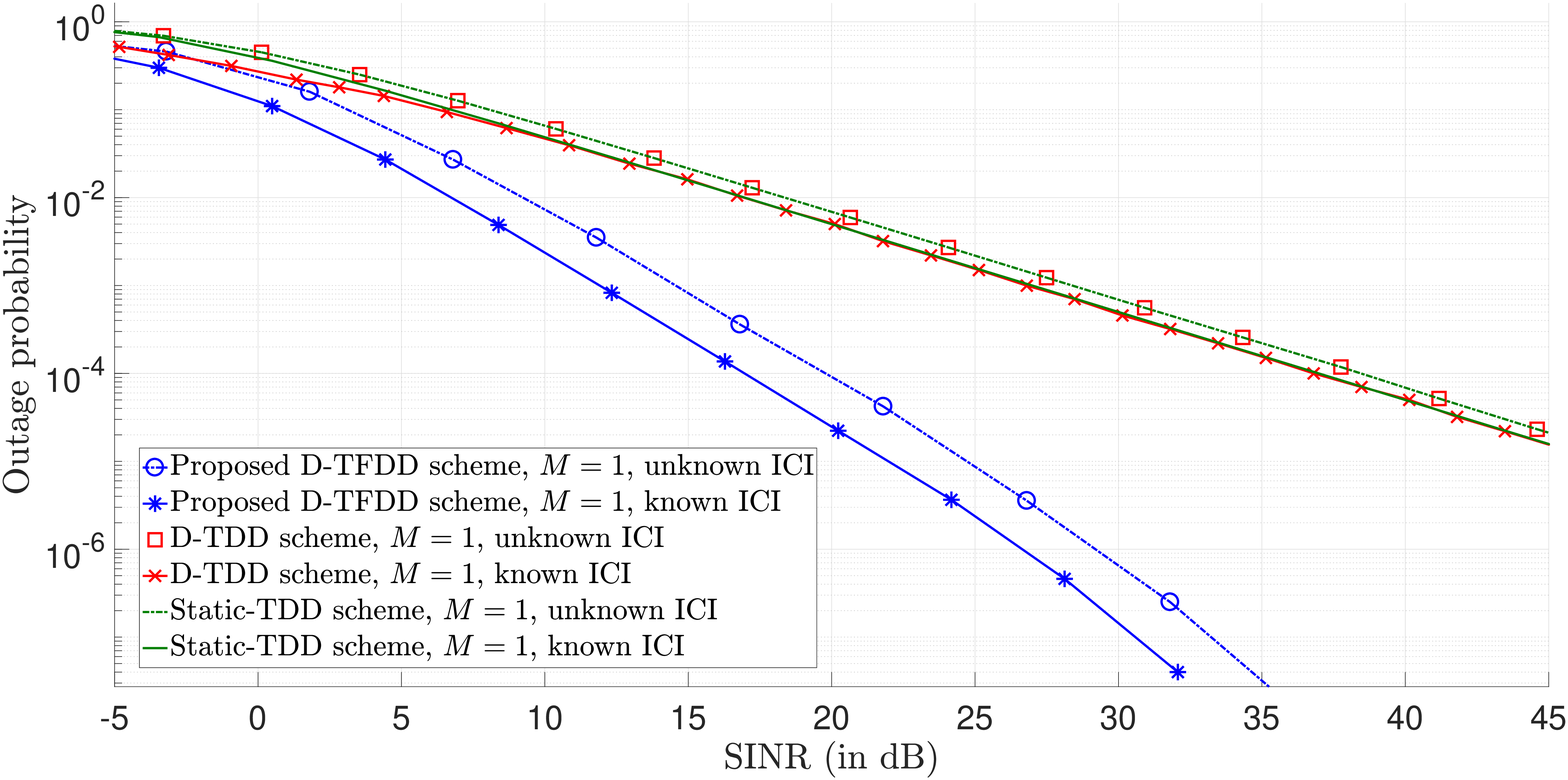}
\caption{Outage probability of the proposed  D-TFDD schemes. Local-CSI and Full-CSI labels highlight that the corresponding schemes are without and with ICI knowledge, respectively.}
\label{Outage_Discrete_1}
\vspace*{-3mm}
\end{figure}

In Fig.~\ref{Outage_Discrete_1}, we illustrate the outage probabilities of the proposed  D-TFDD schemes for unknown ICI and its upper bound, the D-TFDD for known ICI, as well as the benchmark schemes as a function of the  SINR for $M =1$, $\mu=\frac{1}{2}$, where $R$ is set to $R = 1$ bits/symb. As predicted by Theorems~\ref{Theo_PR_Outage_G} and \ref{Theo_PR_Outage_L}, Fig.~\ref{Outage_Discrete_1} shows that the  proposed D-TFDD schemes for unknown and known ICI achieve double the diversity gain compared to the benchmark schemes. Intuitively, the doubling of the diversity gain occurs since the  proposed D-TFDD schemes can select for transmission between two independent channels, in each time slot, compared to the existing D-TDD and D-FDD system, which can select between two dependent channels, in each time slot. The doubling of the diversity gain leads to very large performance gains in terms of SINR. For example, SINR gains of  $10 \; \textrm{dB}$  and $15$ dB can be achieved for outage probabilities of $10^{-2}$ and $10^{-3}$, respectively. On the other hand, the proposed  D-TFDD schemes for unknown ICI has around 3 dB penalty loss compared its   upper bound, the proposed  D-TFDD schemes for known ICI. This example shows  the large performance gains of the proposed  D-TFDD scheme with unknown ICI compared to existing  D-TDD  and/or static-TDD schemes.

\subsubsection{Throughput Region}\label{Sec-RTR}
\begin{figure*}
\vspace*{-2mm}
\centering\includegraphics[width=6in]{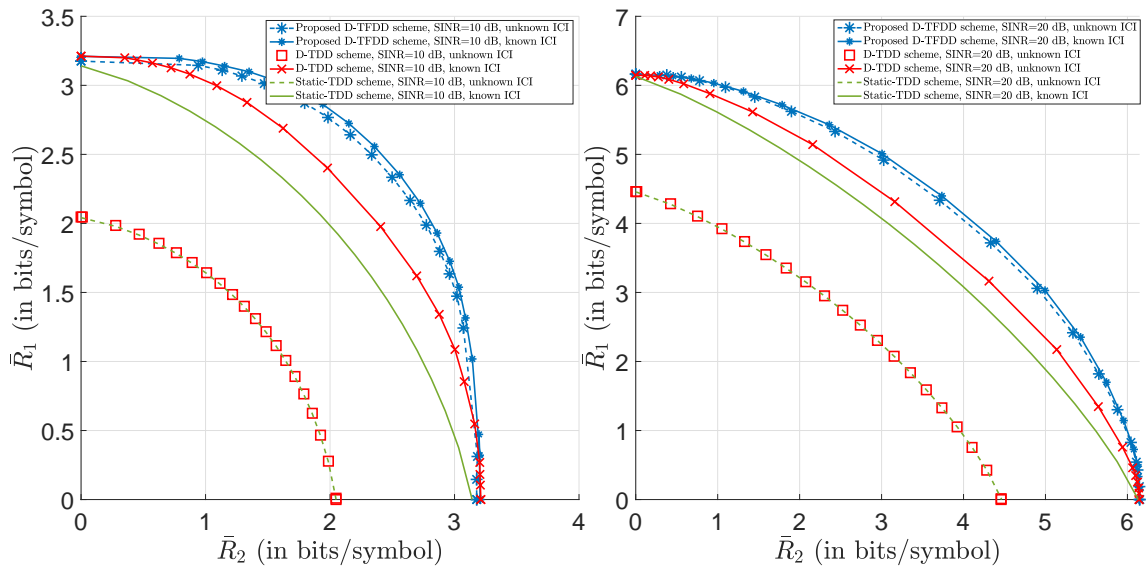}
\caption{Throughput regions of the proposed D-TFDD schemes: left for SINR=10 dB, and right for SINR=20 dB.}
\label{RateRegion_700m}
\vspace*{-3mm}
\end{figure*}

For the example in Fig.~\ref{RateRegion_700m} users have ominidirectional antenna with unity gain and the BS has a directional antenna with gain of $16$ dBi. The power at the users is set to $24$ dBm and the power at BS is set to  $46$ dBm, and we use the proposed scheme with Type 1 and Type 2 for this example.  The distances between U1 and BS, as well as BS and U2, are assumed to be fixed and set to $700$m. The noise figure of BS and U2 are set to $2$ dB and $7$ dB, respectively. The above parameters reflect the parameters used in practice. In addition, the mean power of the channel gains of the BS-U1 and BS-U2 channels are calculated using the standard path-loss model as \cite{6847175,7105650,7463025}
\begin{eqnarray}
E\{|h_k(t)|^2\} = \left(\frac{c}{{4\pi {f_c}}}\right)^2d_k^{ - \beta }\ \textrm{, for } k\in\{1,2\},
\end{eqnarray}
where $c$ is the speed of light, $f_c$ is the carrier frequency, $d_k$ is the distance between the transmitter and the receiver of link $k$, and $\beta$ is the path loss exponent. Moreover, the carrier frequency is set to $f_c=1.9$ GHz, and  we assume $\beta=3.6$ for the BS-U1 and BS-U2 channels. 

In Fig.~\ref{RateRegion_700m}, we show the throughput region achieved with the proposed  D-TFDD schemes for unknown and known ICI with $M=1$, for two different scenarios, one with high interference, SINR=10 dB, and the other one with low interference, SINR=20 dB. Furthermore, we show the throughput regions achieved with the benchmark schemes. For the proposed and the benchmark schemes the value of $R$  is optimized numerically for a given $\mu$ such that the throughput is maximized. As can be seen  from Fig.~\ref{RateRegion_700m}, the proposed D-TDD scheme without ICI knowledge achieves almost the exact throughput region as its upper bound achieved with D-TFDD with full ICI knowledge for SINR=20 dB.  Also, in the relatively high interference region, i.e., SINR=10 dB, the proposed D-TFDD scheme without ICI knowledge achieves a throughput region which is very close to its upper bound achieved with the D-TFDD scheme with ICI knowledge. On the other hand, the throughput that the proposed D-TFDD scheme for unknown ICI  achieves is close or higher  than the throughput  achieved  with the   benchmark D-TDD  scheme with ICI knowledge, which is an interesting result since the proposed scheme without ICI knowledge wastes only two  time slots  compared to the $K$  time slots that the  D-TDD and D-FDD schemes with ICI knowledge waste. More importantly, the gains that the proposed D-TFDD scheme for unknown ICI achieves compared to the benchmark schemes without ICI knowledge are considerable. For example, the proposed  D-TFDD scheme for unknown ICI  has a BS-U1 throughput gain of  about $66\%$, $80\%$ and $100\%$, for SINR=10 dB, and about $17\%$, $18\%$ and $38\%$, for SINR=20 dB, compared to the existing D-TDD and to the static-TDD schemes without ICI knowledge for a BS-U2 throughput of 2, 3, and 4  bits/symb, respectively.

\subsubsection{Sum Throughput}\label{Sec-RTR1}
\begin{figure}[t]
\vspace*{-2mm}
\centering\includegraphics[width=3.4in]{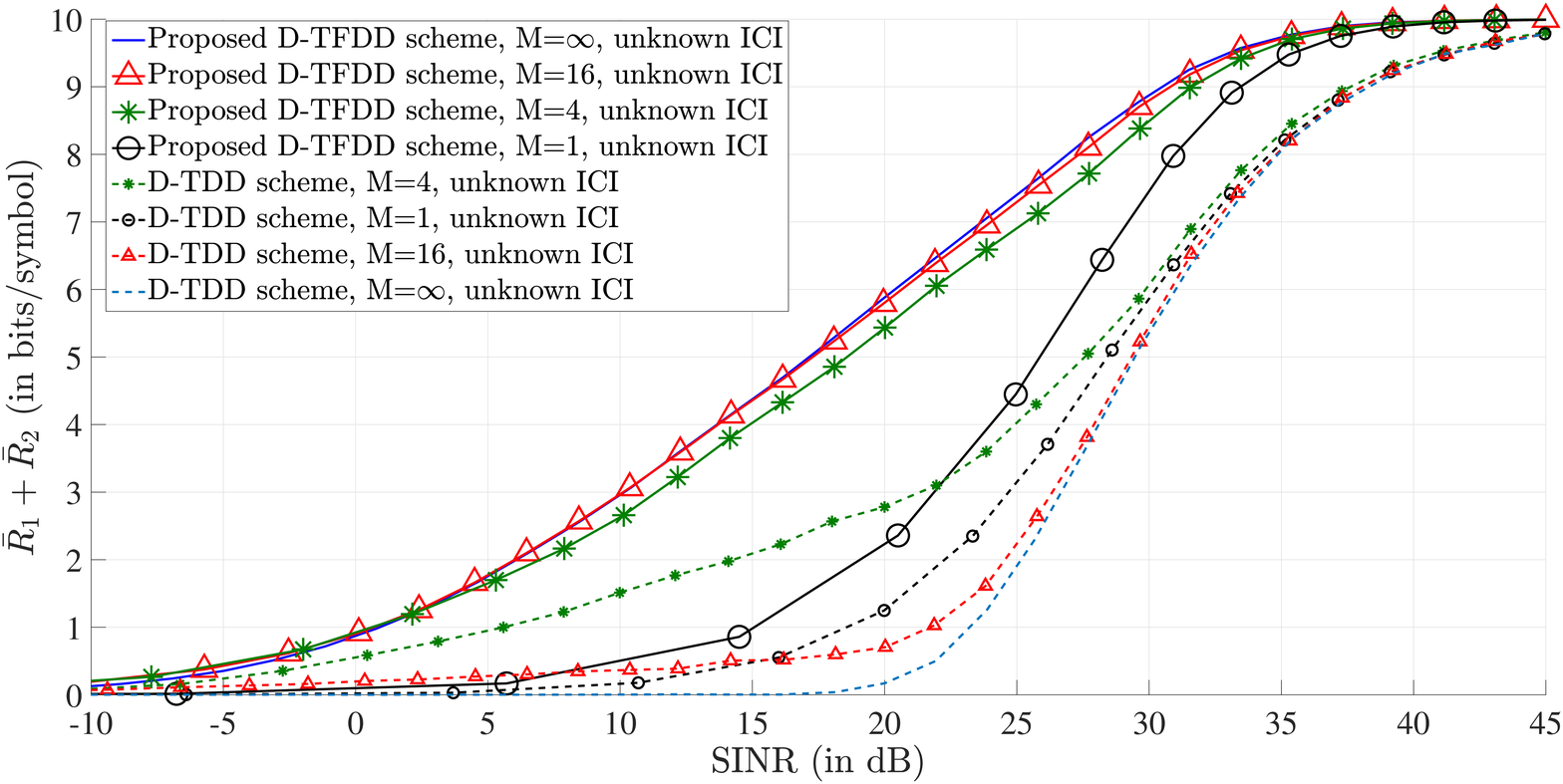}
\caption{Throughputs vs. SINR of the proposed  D-TFDD schemes with and without ICI knowledge with different discrete-rates quantization level.}
\label{discrete_rate_SNR}
\vspace*{-3mm}
\end{figure}

In Fig.~\ref{discrete_rate_SNR}, we illustrate the sum of the BS-U1 and BS-U2 throughputs achieved with the  proposed D-TFDD scheme for unknown ICI  and the  static-TDD as a function of the SINR for $M =1, 4, 16 , \infty$, where $R$ is set to $R = 10/M$ bits/symb. From  Fig.~\ref{discrete_rate_SNR} we can see that  by increasing  $M$ from 1 to 4 in the proposed D-TFDD scheme for unknown  ICI, we can gain more than 10 dBs in  SINR for  around $4$ bits/symb sum throughput. Whereas, by increasing  $M$ from 4 to 16 we can gain an additional 1 dB in  SINR for  around $4$ bits/symb of the sum throughput. Moreover, the proposed D-TFDD scheme for unknown ICI achieves substantial gains compared to the static-TDD. For example, about 3 dB and  10 dB SINR gain is achieved for $M=1$ and $M=4$, respectively, for  around $4$ bits/symb sum throughput. Finally, Fig.~\ref{discrete_rate_SNR} shows that the proposed scheme for $M=4$ performs very close to the case when  $M=\infty$. Note that in Fig.~\ref{discrete_rate_SNR}, the throughputs saturate since as a function of $M$, the transmission  rates are $10/M$ bits/symb. Hence, the maximum possible transmission  rate  is $10$ bits/symb, even if the channels are error-free, which is the case in the high SINR.

The above numerical examples show that the proposed D-TFDD scheme provides 
 double the diversity gain  compared to existing TDD/FDD schemes, which improves that reliability of the communication. Moreover, since the proposed D-TFDD scheme works in a distributed fashion, it does not need any   coordination of the BSs.
Another strength is that is does not require ICI estimation, which makes it practical for implementation. Finally, the proposed D-TFDD scheme
  fits well into the scope of 5G. On the other hand, a weakness of the proposed D-TFDD scheme is the requirement of local CSI of the U1-BS and U2-BS channels at the BS, which entails signalling overhead.

\section{Conclusion}\label{Sec-Conc}

In this paper, we proposed  a distributed D-TFDD scheme   for unknown ICI. Using the proposed D-TFDD scheme, in a given frequency band, the BS   adaptively    selects to either communicate with U1 or   with U2 in a given time slot based on the qualities of the  BS-U1 and BS-U2 channels  without  ICI knowledge such that the BS-U1 and BS-U2 throughput region is maximized. We have shown  that the proposed D-TFDD scheme  provides significant throughput and outage probability gains compared to the conventional static-TDD scheme, as well as to the D-TDD and D-FDD schemes. Moreover, we observed the the proposed D-TFDD scheme doubles the diversity gain on both the BS-U1 and BS-U2 channels  compared to existing duplexing schemes, even when the ICI is unknown, which leads to very large performance gains. 

\appendix 
\subsection{Proof of Theorem~\ref{Theo_PR}}\label{app_PR}
  Constraints C1, C2 and C3 in (\ref{eq_op_PR}) make the problem non-convex. To solve (\ref{eq_op_PR}), we first we relax these constraints to
$0 \leq  q_1^m(t) \leq 1$,
$0 \leq  q_2^l(t) \leq 1$,
and $0 \leq  \sum_{m=1}^M q_1^m(t)+\sum_{l=1}^L q_2^l(t) \leq 1$, thereby making the relaxed problem  convex. The solutions of the relaxed convex problem is then shown to be such that $q_1^m(t)$ and $q_2^l(t)$  take the limiting values 0 or 1, and not the values between 0 and 1. As a result, the relaxed convex problem is equivalent to the original problem. To solve the relaxed problem, we use the Lagrangian. Thereby, we obtain
\begin{align}\label{eq_op_PR_append}
&\begin{array}{ll}
{\cal L} =& - \lim\limits_{T\to\infty}\frac 1T \sum_{t=1}^T \sum_{m=1}^M \mu R_1^m    q_1^m(t) O_1^m  (t)\\
& - \lim\limits_{T\to\infty}\frac 1T \sum_{t=1}^T \sum_{l=1}^L (1-\mu)R_2^l   q_2^l(t) O_2^l  \\
&- { \sum_{m=1}^M {\lambda _1^m(t)} q_1^m(t)} - \left(1 - {\sum_{m=1}^M {\lambda _2^m(t)}q_1^m(t)}\right)
\\
&-{ \sum_{l=1}^L {\lambda _3^l(t)} q_2^l(t)}- \left(1 - {\sum_{l=1}^L {\lambda _4^l(t)}q_2^l(t)}\right) \\\
& -{\lambda _5(t)}\left({\sum_{m=1}^M q_1^m(t)} + \sum_{l=1}^L{ q_2^l(t)}\right) 
\end{array}\nonumber\\[-0.5em]
& \quad\quad\,\,\,\,\,- {\lambda _6(t)}\bigg(1 - \sum_{m=1}^M{ q_1^m(t)} - \sum_{l=1}^L{ q_2^l(t)}\bigg), 
\end{align}
where $\lambda _1^m(t)\geq0$, $\lambda _2^m(t)\geq0$, $\lambda _3^l(t)\geq0$, $\lambda _4^l(t)\geq0$, $\lambda _5(t)\geq0$, and $\lambda_6(t)\geq0$, $\forall m,l,i$, are the Lagrangian multipliers. Next, we rewrite (\ref{eq_op_PR_append}) as
\begin{align}\label{eq_oprew_PR_append}
{\cal L} &=  -\lim\limits_{T\to\infty}\frac 1T \sum_{t=1}^T \mu  q_{1}(t)  {\underset{m }{\textrm{max} }}\{R_1^m O_1^m (t)\}\\
&\begin{array}{ll}
&- \lim\limits_{T\to\infty}\frac 1T \sum_{t=1}^T  (1-\mu)  q_{2}(t) {\underset{l }{\textrm{max} }}\{R_2^l O_2^l (t)\}  \\
&- { \sum_{m=1}^M {\lambda _1^m(t)} q_1^m(t)} - \left(1 - {\sum_{m=1}^M {\lambda _2^m(t)}q_1^m(t)}\right)\\
&- { \sum_{l=1}^L {\lambda _3^l(t)} q_2^l(t)}- \left(1 - {\sum_{l=1}^L {\lambda _4^l(t)}q_2^l(t)}\right) \\
&  -{\lambda _5(t)}\left({  q_{1}(t)} + {q_{2}(t)}\right) - {\lambda _6(t)}\left(1 - {  q_{1}(t)} -{  q_{2}(t)}\right). 
\end{array}\nonumber
\end{align}

Now, using (\ref{eq_oprew_PR_append}) and defining $-\lambda_{5}(t)+\lambda_{6}(t)\triangleq\beta(t)$, we can find the optimal state-selection variables $q_1^m(t)$ and $q_2^l(t)$ as follows. The conditions which maximize (\ref{eq_op_PR_append}), in the cases when the transmit node on the BS-U1 channel transmits with $R_1^m$ and the transmit node on the BS-U2 channel is silent, are
\begin{align}\label{eq_PR_power_q1}
&[ \mu {\underset{m }{\textrm{max} }}\{R_1^m O_1^m (t)\}  - \beta(t)  > 0 ] \;\nonumber\\
\textrm{and} &\; [ (1 - \mu ) {\underset{l }{\textrm{max} }}\{R_2^l O_2^l (t)\}  - \beta(t)  < 0 ].
\end{align}
On the other hand, the conditions for maximizing (\ref{eq_oprew_PR_append}) in the case when the transmit node on the BS-U2 channel transmits with $R_2^l$ and the transmit node on the BS-U1 channel is silent, are
\begin{align}\label{eq_PR_power_q2}
&[ \mu {\underset{m }{\textrm{max} }}\{R_1^m O_1^m (t)\}  - \beta(t)  < 0 ] \;\nonumber\\ \textrm{and} &\; [ (1 - \mu ) {\underset{l }{\textrm{max} }}\{R_2^l O_2^l (t)\}  - \beta(t)  > 0 ].
\end{align}
In (\ref{eq_PR_power_q1}) and (\ref{eq_PR_power_q2}), we can substitute $\mu R_1^m O_1^m (t) $  with $\Lambda_1^m(t)$ and $ (1 - \mu ) R_2^l O_2^l (t) $  with $\Lambda_2^l(t)$, and thereby obtain $q_1^m(t)$ and $q_2^l(t)$ as in (\ref{eq_scheme_AP_G}). This completes the proof.

\subsection{Proof of Theorem~\ref{Theo_PR_Outage_G}}\label{app_PR_Outage_G} 
 
  In time slot $t$, an outage occurs  if the BS-U1 channel is selected for transmission and the BS-U1 channel is too weak to support the rate $R_0$, i.e., $q_1^1(t)=1$ and $O_{1}^1(t)=0$, or if the BS-U2 channel is selected to transmit and the BS-U2 channel is too weak to support the rate $R_0$, i.e., $q_2^1(t)=1$ and $O_{2}^1(t)=0$, or if both the BS-U1 the BS-U2 channels   are not selected for transmission in time slot $t$, i.e., if $q_1^1(t)=q_2^1(t)=0$, since in that case the time slot $t$ is wasted. Hence,  the outage probability  $P_{\rm out}$ can be obtained as
\begin{align}\label{eq_out12_G}
P_{\rm out}&={\rm Pr}\Big\{[q_1^1(t)=1 \textrm { AND } O_{1}^1(t)=0]\nonumber\\
& \quad \quad \quad \textrm{ OR } [q_2^1(t)=1 \textrm { AND } O_{2}^1(t)=0] \nonumber\\ 
& \quad \quad \quad  \textrm{ OR } [q_1^1(t)=q_2^1(t)=0]\Big\}\nonumber\\
& \stackrel{(a)}{=}{\rm Pr}\big\{ q_1^1(t)=1 \textrm { AND } O_{1}^1(t)=0 \big\} \nonumber\\
&+ {\rm Pr}\big\{ q_2^1(t)=1 \textrm { AND } O_{2}^1(t)=0 \big\}\nonumber\\ 
&+{\rm Pr}\big\{q_1^1(t)=q_2^1(t)=0\big\},
\end{align}
where $(a)$ follows since the events $q_1^1(t)=1$ and $q_1^1(t)=0$, and also the events $q_2^1(t)=1$ and $q_2^1(t)=0$ are mutually exclusive. Since $\mu=\frac{1}{2}$, $\Lambda_1^1(t)$  and $\Lambda_2^1(t)$ in (\ref{eq_PR_1}) and (\ref{eq_PR_2}) simplify to
\begin{align}
\Lambda_k^1(t) &=\frac{1}{2}  R_0 O_k^1 (t)  ,\; k\in\{1,2\}.\label{eq_PR_1a}
\end{align}
Now, inserting $\Lambda_1^1(t)$  and $\Lambda_2^1(t)$ from (\ref{eq_PR_1a}) into (\ref{eq_scheme_AP_G}), we obtain that $q_1^1(t)=1$ if $O_1^1 (t)\geq O_2^1 (t)$ and $O_1^1 (t)>0$, which means that $q_1^1(t)=1$ occurs if $O_1^1 (t)=1$. Hence, the event $q_1^1(t)=1$ and $O_1^1 (t)=0$ is an impossible event, thereby leading to ${\rm Pr}\big\{ q_2^1(t)=1 \textrm { AND } O_2^1(t)=0 \big\}=0$ in (\ref{eq_out12_G}). Similarly, we can conclude that ${\rm Pr}\big\{ q_1^1(t)=1 \textrm { AND } O_1^1(t)=0 \big\}=0$ in (\ref{eq_out12_G}). Next, we obtain that $q_1^1(t)=q_2^1(t)=0$ occurs iff $O_1^1 (t)= O_2^1 (t)=0$ holds, thereby leading to ${\rm Pr}\big\{q_1^1(t)=q_2^1(t)=0\big\}={\rm Pr}\big\{O_1^1 (t)=0 \textrm{ AND } O_2^1 (t)=0\big\}$ in (\ref{eq_out12_G}). Inserting this into (\ref{eq_out12_G}), we obtain  
\begin{align}\label{eq_out1_2a}
P_{\rm out}&={\rm Pr}\big\{O_1^1 (t)=0 \textrm{ AND } O_2^1 (t)=0\big\} \\
&={\rm Pr}\Bigg\{\log_2\left(1 + P\frac{|\gamma_{1}(t)|^2}{ 1+\gamma _{I1}(t)}\right)< R_0 \nonumber\\
&\textrm{ AND } \log_2\left(1 + P\frac{|\gamma_{2}(t)|^2}{ 1+\gamma _{I2}(t)}\right)< R_0\Bigg\}\nonumber\\
& ={\rm Pr}\left\{ \frac{|\gamma_{1}(t)|^2}{ 1+\gamma _{I1}(t)}<\gamma_{\rm th} \textrm{ AND } \frac{|\gamma_{2}(t)|^2}{ 1+\gamma _{I2}(t)}<\gamma_{\rm th} \right\},\nonumber
\end{align} 
where $\gamma_{\rm th}=\frac{2^{R_0}-1}{P} $.  The variables $\gamma_{1}(t)$ and $\gamma_{2}(t)$ have identical and independent exponential distributions with PDFs denoted by $f_{\gamma_{1}}(\gamma_{1})$ and $f_{\gamma_{2}}(\gamma_{2})$, respectively, both with mean $\Omega_0=E\{\frac{|h_{1}(t)|^2}{\sigma_1^2}\}=E\{\frac{|h_{2}(t)|^2}{\sigma_2^2 }\}$. On the other hand, the variables $\gamma_{I1}(t)$ and $\gamma_{I2}(t)$ have identical yet dependent exponential distributions with joint PDF denoted by $f_{\gamma_{I1},\gamma_{I2}}\big(Z_1,Z_2\big)$.
As a result, (\ref{eq_out1_2a}) can be obtained as
\begin{align}
P_{\rm out}= &\int\limits_{0}^{\infty} \int\limits_{0}^{\infty} \int\limits_{0}^{\gamma_{\rm th} \big(1+Z_1\big)} {} \int\limits_{0}^{\gamma_{\rm th} \big(1+Z_2\big)}  & \hspace{-5mm}f_{\gamma_{1},\gamma_{2},\gamma_{I1},\gamma_{I2}}\big(\gamma_{1},\gamma_{2},Z_1,Z_2\big) \nonumber\\
& \times d{\gamma_{1}} d{\gamma_{2}} d{Z_1} d{Z_2}  .\label{Out_U5GG1}
\end{align}
We can rewrite $f_{\gamma_{1},\gamma_{2},\gamma_{I1},\gamma_{I2}}\big(\gamma_{1},\gamma_{2},Z_1,Z_2\big)$ in (\ref{Out_U5GG1}) as 
\begin{align}\label{Out_U5GG1123}
f_{\gamma_{1},\gamma_{2},\gamma_{I1},\gamma_{I2}}&\big(\gamma_{1},\gamma_{2},Z_1,Z_2\big)
=f_{\gamma_{1}|\gamma_{I1},\gamma_{I2}}\big(\gamma_{1}\big)\nonumber\\
& \times f_{\gamma_{2}|\gamma_{I1},\gamma_{I2}}\big(\gamma_{2}\big) f_{\gamma_{I1},\gamma_{I2}}\big(Z_1,Z_2\big),
\end{align}
since $\gamma_{1}$ and $\gamma_{2}$ have independent distributions. By substituting the PDFs of $\gamma_{1}$ and $\gamma_{2}$ into (\ref{Out_U5GG1123}), then inserting  (\ref{Out_U5GG1123}) into (\ref{Out_U5GG1}), we can obtain (\ref{Out_U5GG1}) as
\begin{align}
P_{\rm out}  =  \int\limits_{0}^{\infty} \hspace{-0.5mm}& \int\limits_{0}^{\infty}  \frac{1}{\Omega_0} \hspace{-3.5mm}	 \int\limits_{0}^{\gamma_{\rm th} (1+Z_2)} \hspace{-3.5mm}  e^{-(\frac{\gamma_{2}}{\Omega_0})} d{\gamma_{2}}   
\times \frac{1}{\Omega_0}\hspace{-3.5mm}	 \int\limits_{0}^{\gamma_{\rm th} (1+Z_1)}  e^{-(\frac{\gamma_{1}}{\Omega_0})}    d{\gamma_{1}}  \nonumber\\
& \times f_{\gamma_{I1},\gamma_{I2}}(Z_1,Z_2)  d{Z_1} d{Z_2}  .\label{Out_U5GG2}
\end{align}

In (\ref{Out_U5GG2}), by integrating over $\gamma_{1}$ and $\gamma_{2}$, and letting $P \to \infty$ (consequently $\gamma_{\rm th} \to 0$), we obtain
\begin{align}
P_{\rm out} \to \int\limits_{0}^{\infty}  \int\limits_{0}^{\infty} &  
\frac{\gamma_{\rm th} (1+Z_2)}{\Omega_0} \times \frac{\gamma_{\rm th} (1+Z_1)}{\Omega_0} \nonumber\\	
 &\times f_{\gamma_{I1},\gamma_{I2}}(Z_1,Z_2)  d{Z_1} d{Z_2}  \; \textrm{as} \;{P\to\infty} .\label{Out_U5GG3}
\end{align}

Finally, by replacing 
\begin{align}
&E \Big \{  (1+\gamma _{I1}(t)) (1+\gamma _{I2}(t)) \Big \} \\ &=\int\limits_{0}^{\infty}  \int\limits_{0}^ {\infty}    (1+Z_2)\times  (1+Z_1)	\times f_{\gamma_{I1},\gamma_{I2}}(Z_1,Z_2)  d{Z_1} d{Z_2},\nonumber
\end{align}
into (\ref{Out_U5GG3}), the outage is obtained as in (\ref{Out_U6GG}). This completes the proof. 

\subsection{Proof of Proposition~\ref{Theo_PR_L0cal}}\label{app_PR_Local}

The BS-U1 and BS-U2 throughputs are given in (\ref{eq_PR_fr3}) and (\ref{eq_PR_fr4}), respectively. However, since we can not compute $O_{1}^{m}(t)$ and $O_{2}^{l}(t)$, instead of  (\ref{eq_PR_fr3}) and (\ref{eq_PR_fr4}), we  use estimates for  (\ref{eq_PR_fr3}) and (\ref{eq_PR_fr4}), given by
\begin{align}
\bar R_{1,e} &=\lim_{T\to\infty}\frac 1T \sum_{t=1}^T \sum_{m=1}^M R_1^m   q_1^m(t) O_{1,e}^m  (t),\label{eq_PR_fr3123}
\\
\bar R_{2,e} &=\lim_{T\to\infty}\frac 1T \sum_{t=1}^T \sum_{l=1}^L R_2^l   q_2^l(t) O_{2,e}^l  (t).\label{eq_PR_fr4123}
\end{align}
The accuracy of the estimates $\bar R_{1,e}$ and $\bar R_{2,e}$, depends on the following expressions
\begin{align}
\delta_1=&\lim_{T\to\infty}\frac 1T \sum_{t=1}^T q_1^m(t) \bigg ( O_{1}^{m}(t)-O_{1,e}^{m}(t)  \bigg  )^2. \label{eq_PR_fr334}\\
\delta_2=&\lim_{T\to\infty}\frac 1T \sum_{t=1}^T q_2^l(t) \bigg  ( O_{2}^{l}(t)-O_{2,e}^{l}(t)  \bigg  )^2.\label{eq_PR_fr434}
\end{align}
Hence, (\ref{eq_PR_fr334}) and (\ref{eq_PR_fr434}) express the average of the difference  squared between the outages when the ICI is known and the estimation of the outages when the ICI is unknown. The smaller (\ref{eq_PR_fr334}) and (\ref{eq_PR_fr434}) are, the more accurate the estimates $\bar R_{1,e}$ and  $\bar R_{2,e}$ become. In fact, when $\delta_1 \to 0$, and $\delta_2 \to 0$, $\bar R_{1,e} \to \bar R_{1}$ and  $\bar R_{2,e} \to \bar R_{2}$.

Now, if $\delta_1 < \epsilon$ and $\delta_2 < \epsilon$ hold, the constants $\gamma _{I1}$ and $\gamma _{I2}$ that  maximize the  estimated BS-U1 and BS-U2 throughput region, defined in (\ref{eq_PR_fr3123}) and (\ref{eq_PR_fr4123}), can be found from the following maximization problem 
\begin{align}\label{eq_op_PRlt}
& {\underset{\gamma _{I1}, \gamma _{I2}.\;}{\textrm{Maximize: }}}\;  \lim_{T\to\infty}\frac 1T \sum_{t=1}^T  \mu R_1^{m^*}   q_1^{m^*}(t) O_{1,e}^{m^*}(t) \nonumber\\
 &\qquad \qquad+ \lim_{T\to\infty}\frac 1T \sum_{t=1}^T \big(1 - {\mu} \big) R_2^{l^*}   q_2^{l^*}(t) O_{2,e}^{l^*}(t) \nonumber\\
 &  {\rm{Subject\;\;  to \; :}} \nonumber\\
& \qquad {\rm C1:} \;    \lim_{T\to\infty}\frac 1T \sum_{t=1}^T q_1^{m^*}(t)\bigg ( O_{1}^{m^*}(t)-O_{1,e}^{m^*}(t)  \bigg  )^2\leq \epsilon, \nonumber\\ 
& \qquad {\rm C2:} \;    \lim_{T\to\infty}\frac 1T \sum_{t=1}^T  q_2^{l^*}(t)\bigg  ( O_{2}^{l^*}(t)-O_{2,e}^{l^*}(t)  \bigg  )^2 \leq  \epsilon,
\end{align}
where $m^*=\arg {\underset{m }{\textrm{max} }}\{R_1^m O_{1,e}^m (t)\}$ and $l^*=\arg{\underset{l }{\textrm{max} }}\{R_2^l O_{2,e}^l (t)\}$. 

By applying the Lagrangian function on (\ref{eq_op_PRlt}), we obtain
\begin{align}\label{eq_op_PR_appendl}
&\begin{array}{ll}
{\cal L} =& - \lim\limits_{T\to\infty}\frac 1T \sum_{t=1}^T  \mu R_1^{m^*}    q_1^{m^*}(t) O_{1,e}^{m^*}(t) \nonumber\\
&- \lim\limits_{T\to\infty}\frac 1T \sum_{t=1}^T  (1-\mu)R_2^{l^*}   q_2^{l^*}(t) O_{2,e}^{l^*}(t)  \\
 &+\lim\limits_{T\to\infty}\frac 1T \sum_{t=1}^T  \chi_1 q_1^{m^*}(t) \bigg ( O_{1}^{m^*}(t)-O_{1,e}^{m^*}(t) \bigg )^2
\end{array}\nonumber\\[-0.5em]
&\qquad\quad+ \lim\limits_{T\to\infty}\frac 1T \sum_{t=1}^T  \chi_2 q_2^{l^*}(t)\bigg ( O_{2}^{l^*}(t)-O_{2,e}^{l^*}(t) \bigg)^2,
\end{align}
where $\chi_1 \geq 0$ and $\chi_2 \geq 0$ are the Lagrangian multipliers,  found such that C1 and C2 in (\ref{eq_op_PRlt}) hold. By differentiating ${\cal L}$ in (\ref{eq_op_PR_appendl}) with respect to $\gamma _{I1}$ and $\gamma _{I2}$, and equivalenting the results to zero, we obtain
\begin{align}\label{gamaU_l11}
&\lim\limits_{T\to\infty}\frac 1T \sum_{t=1}^T  \Bigg [ \frac{ - P_1 \gamma_{1}(t) \delta_1^{m^*}(t) q_1^{m^*}(t) }{\ln(2) \Big (1+\gamma _{I1}+P_1 \gamma _{1}(t)\Big) \Big (1+\gamma _{I1}\Big)} \nonumber\\
& \quad \times \Big ( -\mu R_1^{m^*}   + 2\chi_1 \big [ O_{1,e}^{m^*}(t)-O_{1}^{m^*}(t) \big] \Big) \Bigg ]=0 ,
\end{align}
\begin{align}\label{gamaD_l11}
&\lim\limits_{T\to\infty}\frac 1T \sum_{t=1}^T \Bigg [ \frac{ - P_2 \gamma_{2}(t) \delta_2^{l^*}(t) q_2^{l^*}(t)}{\ln(2) \Big (1+\gamma _{I2}+P_2 \gamma _{2}(t)\Big) \Big (1+\gamma _{I2}\Big)}\nonumber\\
& \quad \times \Big ( -(1-\mu) R_2^{l^*}   + 2\chi_2 \big [ O_{2,e}^{l^*}(t)-O_{2}^{l^*}(t) \big] \Big)\Bigg ]=0.
\end{align} 
Due to the law of large numbers, (\ref{gamaU_l11}) and (\ref{gamaD_l11}) can be written as 
\begin{align}\label{gamaU_l11a}
&E \Bigg [ \frac{ - P_1 \gamma_{1}(t) \delta_1^{m^*}(t) q_1^{m^*}(t)}{\ln(2) \Big (1+\gamma _{I1}+P_1 \gamma _{1}(t)\Big) \Big (1+\gamma _{I1}\Big)} \nonumber\\
& \quad  \times \Big ( -\mu R_1^{m^*}   + 2\chi_1  \big [ O_{1,e}^{m^*}(t)-O_{1}^{m^*}(t) \big] \Big) \Bigg ] =0 ,
\end{align}
\begin{align}\label{gamaD_l11a}
&E \Bigg [\frac{ - P_2 \gamma_{2}(t) \delta_2^{l^*}(t) q_2^{l^*}(t)}{\ln(2) \Big (1+\gamma _{I2}+P_2 \gamma _{2}(t)\Big) \Big (1+\gamma _{I2}\Big)} \nonumber\\
& \quad  \times \Big ( -(1-\mu) R_2^{l^*}   + 2\chi_2 \big [ O_{2,e}^{l^*}(t)-O_{2}^{l^*}(t) \big] \Big) \Bigg]=0.
\end{align} 

Calculating the constants $\gamma _{I1}$ and $\gamma _{I2}$ from (\ref{gamaU_l11a}) and (\ref{gamaD_l11a}) requires the derivation of the above expectations. Instead, we use a more practical approach,  where the constants $\gamma _{I1}$ and $\gamma _{I2}$ are  estimated as  $\gamma_{I1}^e(t)$ and $\gamma_{I2}^e(t)$ in time slot $t$.  To this end, we apply the gradient descent method \cite{Boyd_CO} on (\ref{gamaU_l11}) and (\ref{gamaD_l11}) to obtain $\gamma_{I1}^e(t)$ and $\gamma_{I2}^e(t)$ as in (\ref{eq_z1_Rsr1}) and (\ref{eq_z1_Rsr2}), where $\delta_{k}( t)$ for $k\in\{1,2\}$ is an adaptive step size which controls the speed of convergence of $\gamma_{Ik}^e(t)$ to $\gamma_{Ik}$, for $k\in\{1,2\}$, which can be some properly chosen monotonically decaying function of $t$ with $\delta_{k}( 1)<1$. Note that  $\lim\limits_{t\to\infty} \gamma_{I1}^e(t) = \gamma_{I1}$ and $\lim\limits_{t\to\infty} \gamma_{I2}^e(t) = \gamma_{I2}$. This completes the proof.
 
\subsection{Proof of Theorem~\ref{Theo_PR_Outage_L}}\label{app_PR_Outage_L} 
 In time slot $t$, an outage occurs  if the BS-U1 channel is selected for transmission and the BS-U1   channel is too weak to support the rate $R_0$, i.e., $q_1^1(t)=1$ and $O_{1}^1(t)=0$, or if the BS-U2 channel is selected to transmission and the BS-U2 channel is too weak to support the rate $R_0$, i.e., $q_2^1(t)=1$ and $O_{2}^1(t)=0$, or if both the BS-U1 and BS-U2 channels are not selected for transmission in time slot $t$, i.e., if $q_1^1(t)=q_2^1(t)=0$, since in that case the time slot $t$ is wasted. Assuming that $\gamma_{I1}^e(t)$ and $\gamma_{I2}^e(t)$ have converged to their steady states  given by $\mathop {\lim }\limits_{t \to \infty } \gamma_{I1}^e(t) =\gamma_{I1}$ and $\mathop {\lim }\limits_{t \to \infty } \gamma_{I2}^e(t) =\gamma_{I2}$, the outage probability, $P_{\rm out}$, can be found as
\begin{align}\label{eq_out12}
P_{\rm out}&={\rm Pr}\Big\{[q_1^1(t)=1 \textrm { AND } O_{1}^1(t)=0] \nonumber\\ 
& \quad \quad \quad \textrm{ OR } [q_2^1(t)=1 \textrm { AND } O_{2}^1(t)=0] \nonumber\\ 
& \quad \quad \quad  \textrm{ OR } [q_1^1(t)=q_2^1(t)=0]\Big\}\nonumber\\
& \stackrel{(a)}{=}{\rm Pr}\big\{ q_1^1(t)=1 \textrm { AND } O_{1}^1(t)=0 \big\} \nonumber\\
&+ {\rm Pr}\big\{ q_2^1(t)=1 \textrm { AND } O_{2}^1(t)=0 \big\}\nonumber\\ 
&+{\rm Pr}\big\{q_1^1(t)=q_2^1(t)=0\big\},
\end{align}
where $(a)$ follows since the events $q_1^1(t)=1$ and $q_1^1(t)=0$, and also the events $q_2^1(t)=1$ and $q_2^1(t)=0$ are mutually exclusive. 

We divide (\ref{eq_out12}) into three events; BS-U1  communication event, $[q_1^1(t)=1 \textrm { AND } O_{1}^1(t)=0] $,    BS-U2 communication event
$[q_2^1(t)=1 \textrm { AND } O_{2}^1(t)=0]$, and the silent event  $[q_1^1(t)=q_2^1(t)=0]$. In the following, we  calculate the probability of these three events.

For the BS-U1  communication event, we have $q_1^1(t)=1$ when either of the two following events occur

- $O_{1,e}^1 (t)=1$ and $O_{2,e}^1(t)=0$. This event occurs when  $\frac{\gamma_{1}(t)}{1+\gamma _{I1}} \geq \gamma_{\rm th}$ and $\frac{\gamma_{2}(t)}{1+\gamma _{I2}}  < \gamma_{\rm th}$, where $\gamma_{\rm th}=\frac{2^{R_0}-1}{P} $.

- $O_{1,e}^1 (t)=O_{2,e}^1(t)=1$ and $\gamma_{U}^e(t) \geq \gamma_{D}^e(t)$. This event occurs when $\frac{\gamma_{1}(t)}{1+\gamma _{I1}} > \gamma_{\rm th}$, $\frac{\gamma_{2}(t)}{1+\gamma _{I2}}  > \gamma_{\rm th}$, and $\frac{\gamma_{1}(t)}{1+\gamma _{I1}} \geq \frac{\gamma_{2}(t)}{1+\gamma _{I2}} $.

On the other hand, the event $O_{1}^1(t)=0$ occurs with the following probability 
\begin{align}\label{Otage_Z}
{\rm Pr}\Big\{O_{1}^1(t)=0 \Big\}= {\rm Pr} \left({\frac{\gamma_{1}(t)}{1+\gamma_{I1}(t)} < \gamma_{\rm th} }\right),
\end{align}
where $\gamma_{I1}(t)$ is the sum of $K$ identical and independent random variables  with mean (identical mean is assumed for simplicity) $\Omega_I$, which has the following PDF
\begin{align}\label{PDF_erl}
f_{\gamma_{I1}}(z)=\frac{z^{K-1} e^{-\frac{z}{\Omega_I}}}{\Omega_I^K (K-1)!}
\end{align}
and the following cumulative distribution function (CDF)
\begin{align}\label{CDF_erl}
F_{\gamma_{I1}}(z)=1- \sum_{n=0}^{K-1} \frac{1}{n!} e^{-\frac{z}{\Omega_I}} {\bigg(\frac{z}{\Omega_I} \bigg)^n}.
\end{align}

In addition, the variables $\gamma_{1}(t)$ and $\gamma_{2}(t)$  have i.i.d. exponential distributions with PDFs, $f_{\gamma_{1} }(\gamma_{1} )$ and $f_{\gamma_{2}}(\gamma_{2})$, respectively, that have mean $\Omega_0=E\{\frac{|h_{1}(t)|^2}{\sigma_1^2 }\}=E\{\frac{|h_{2}(t)|^2}{\sigma_2^2 }\}$.

Using the above,  we can rewrite ${\rm Pr}\Big\{q_1^1(t)=1 \textrm { AND } O_{1}^1(t)=0 \Big\}$ in an integral form as
\begin{align} \label{Out_U1}
&{\rm Pr}\Big\{q_1^1(t)=1 \textrm { AND } O_{1}^1(t)=0 \Big\}\nonumber\\ 
&=    {\int\limits_{\gamma_{\rm th} ({1+\gamma _{I2}})}^{\infty} \hspace{-3mm}  \int\limits_{0}^{\gamma_{\rm th}({1+\gamma _{I1}}) } \hspace{-3mm} {\rm Pr} \hspace{-.5mm} \left( \hspace{-.5mm}{   Z \hspace{-.5mm}> \hspace{-.5mm}\frac{\gamma_{1}}{\gamma_{\rm th} } \hspace{-.5mm}-\hspace{-0.5mm}1\hspace{-.5mm} }\right) \hspace{-.5mm}f_{\gamma_{1}}(\gamma_{1}) f_{\gamma_{2}}(\gamma_{2}) d{\gamma_{2}} d{\gamma_{1}} }\nonumber\\ 
&+ { \int\limits_{\gamma_{\rm th} ({1+\gamma _{I2}})}^{\infty} \hspace{-0mm} \int\limits_{\gamma_{2} \big( \frac{1+\gamma _{I1}}{1+\gamma _{I2}} \big)}^{\infty}  \hspace{-5mm} {\rm Pr} \hspace{-0.5mm} \left({ \hspace{-0.5mm}  Z\hspace{-0.5mm} >\hspace{-0.5mm}\frac{\gamma_{1}}{\gamma_{\rm th} } \hspace{-0.5mm}-\hspace{-0.5mm}1\hspace{-0.5mm} }\right)f_{\gamma_{1}}(\gamma_{1}) f_{\gamma_{2}}(\gamma_{2})   d{\gamma_{1}} d{\gamma_{2}}} \nonumber\\ 
&\mathop  = \limits^{(a)} \int\limits_{\gamma_{\rm th}({1+\gamma _{I1}})}^{\infty} \hspace{-3mm} \int\limits_{0}^{\gamma_{1} \big( \frac{1+\gamma _{I2}}{1+\gamma _{I1}} \big)} \sum_{n=0}^{K-1} \frac{1}{n!} e^{-\left(\frac{[\frac{\gamma_{1}}{\gamma_{\rm th}}-1]}{\Omega_I} \right )}  {\left(\frac{[\frac{\gamma_{1}}{\gamma_{\rm th}}-1]}{\Omega_I} \right)^n} \nonumber\\
& \qquad\times f_{\gamma_{1}}(\gamma_{1}) f_{\gamma_{2}}(\gamma_{2})  d{\gamma_{2}}  d{\gamma_{1}} ,
\end{align}
where (a) follows from (\ref{CDF_erl}). Now,  performing the integration with respect to  $\gamma_{2}$, we obtain
\begin{align} \label{Out_U2}
&{\rm Pr}\Big\{q_1^1(t)=1 \textrm { AND } O_{1}^1(t)=0 \Big\}\nonumber\\ 
&= \frac{\gamma_{\rm th} \Omega_I e^{- (\frac{\gamma_{\rm th}}{\Omega_0})}}{\Omega_0} \int\limits_{\frac{\gamma _{I1}}{\Omega_I}}^{\infty}   \sum_{n=0}^{K-1} \frac{1}{n!} e^{-U'}  {(U')^n} e^{-\left(\frac{\gamma_{\rm th} \Omega_I U'}{\Omega_0 } \right)} \nonumber\\
& \times \left(1-e^{- \left[(\frac{\gamma_{\rm th} }{\Omega_0 }) (1+\Omega_I U') (\frac{1+\gamma _{I2}}{1+\gamma _{I1}} ) \right ]} \right) dU',
\end{align}
where $U'=\frac{ \left(\frac{\gamma_{1}}{\gamma_{\rm th}}-1 \right)}{\Omega_I}$. In (\ref{Out_U2}), when  $P \to \infty$, and consequently $\gamma_{\rm th} \to 0$, we can approximate  $(1-e^{-[(\frac{\gamma_{\rm th} }{\Omega_0 }) (1+\Omega_I U') (\frac{1+\gamma _{I2}}{1+\gamma _{I1}} ) ]})$ by $({[(\frac{\gamma_{\rm th} }{\Omega_0 }) (1+\Omega_I U') (\frac{1+\gamma _{I2}}{1+\gamma _{I1}} ) ]})$,   $e^{-(\frac{\gamma_{\rm th}}{\Omega_0})}$ by 1, and $e^{-(\frac{\gamma_{\rm th} \Omega_I U'}{\Omega_0 })}$ by 1. As a result, (\ref{Out_U2}) can be rewritten as
\begin{align} 
& {\rm Pr}\Big\{q_1^1(t)=1 \textrm { AND } O_{1}^1(t)=0 \Big\}\label{Out_U3}\\ 
&\to \frac{ \gamma_{\rm th}^2 \Omega_I }{\Omega_0^2} \times \frac{1+\gamma _{I2}}{1+\gamma _{I1}}  \hspace{-2mm}\Bigg[\sum_{n=0}^{K-1} \frac{1}{n!} \int\limits_{\frac{\gamma _{I1}}{\Omega_I}}^{\infty} {   e^{-U'}  {(U')^n}   dU'} \nonumber\\
&+ \Omega_I \sum_{n=0}^{K-1} \frac{1}{n!} \int\limits_{\frac{\gamma _{I1}}{\Omega_I}}^{\infty} {   e^{-U'}  {(U')^{n+1}}   dU'}\Bigg] \hspace{-2mm} \;\;\; \textrm{as} \;P \to \infty.\nonumber
\end{align}

By calculating the integrals and the summations in (\ref{Out_U3}), we obtain
\begin{align} \label{Out_U4}
{\rm Pr}\Big\{ \hspace{-0.25mm}q_1^1(t)\hspace{-0.5mm}=\hspace{-0.5mm}1 \hspace{-0.5mm} \textrm { AND } \hspace{-0.5mm} O_{1}^1(t)\hspace{-0.5mm}=\hspace{-0.5mm}0 \hspace{-0.25mm} \Big\} \hspace{-1mm} \to \hspace{-1mm} \frac{ \gamma_{\rm th}^2 {\hat \Omega_{I1}} }{\Omega_0^2} \; \textrm{as} \;P  \hspace{-1mm}\to  \hspace{-1mm}\infty,
\end{align}
where ${\hat \Omega_{I1}}$ is given in (\ref{Otage_U_final}).

For the  BS-U2 communication event, we obtain that $q_2^1(t)=1$ if $O_{2,e}^1 (t)=1$ and $O_{1,e}^1(t)=0$,  or $O_{2,e}^1 (t)=O_{1,e}^1(t)=1$ and $\frac{\gamma_{1}(t)}{1+\gamma _{I1}}<\frac{\gamma_{2}(t)}{1+\gamma _{I2}}$. The  event  $O_{2,e}^1 (t)=1$ and $O_{1,e}^1(t)=0$ occurs when  $\frac{\gamma_{2}(t)}{1+\gamma _{I2}}  > \gamma_{\rm th}$ and $\frac{\gamma_{1}(t)}{1+\gamma _{I1}} < \gamma_{\rm th}$. The event $O_{2,e}^1 (t)=O_{1,e}^1(t)=1$ and $\frac{\gamma_{1}(t)}{1+\gamma _{I1}}<\frac{\gamma_{2}(t)}{1+\gamma _{I2}}$ occurs when $\frac{\gamma_{2}(t)}{1+\gamma _{I2}}  > \gamma_{\rm th}$, $\frac{\gamma_{1}(t)}{1+\gamma _{I1}}  > \gamma_{\rm th}$, and $\frac{\gamma_{1}(t)}{1+\gamma _{I1}}<\frac{\gamma_{2}(t)}{1+\gamma _{I2}}$. Using this, we can derive ${\rm Pr}\Big\{q_2^1(t)=1 \textrm { AND } O_{2}^1(t)=0 \Big\}$ with a similar approach as the calculation of ${\rm Pr}\Big\{q_1^1(t)=1 \textrm { AND } O_{1}^1(t)=0 \Big\}$, which results in 
\begin{align}\label{Out_U44}
  {\rm Pr}\Big\{q_2^1(t) \hspace{-0.5mm}=\hspace{-0.5mm}1 \textrm { AND } O_{2}^1(t)\hspace{-0.5mm}=\hspace{-0.5mm}0 \Big\} \hspace{-1mm}\to \hspace{-1mm} \frac{ \gamma_{\rm th}^2 {\hat \Omega_{I2}} }{\Omega_0^2} \; \textrm{as} \;P  \hspace{-1mm}\to \hspace{-1mm} \infty,
\end{align}
where ${\hat \Omega_{I2}}$ is given in (\ref{Otage_D_final}).

 Finally, for the silent event, we obtain that $q_1^1(t)=q_2^1(t)=0$ occurs iff $O_{1,e}^1 (t)= O_{2,e}^1 (t)=0$ holds, which occurs when $\frac{\gamma_{2}(t)}{1+\gamma _{I2}} < \gamma_{\rm th}$, and $\frac{\gamma_{1}(t)}{1+\gamma _{I1}} < \gamma_{\rm th}$. The probability of the silent event can be calculated by
 \begin{align} \label{Out_U5}
  & {\rm Pr}\big\{q_1^1(t)=q_2^1(t)=0\big\} \\ &=\frac{1}{\Omega_0^2}\int\limits_{0}^{\gamma_{\rm th} ({1+\gamma _{I2}})} {\int\limits_{0}^{\gamma_{\rm th} ({1+\gamma _{I1}})}  e^{-(\frac{\gamma_{1}}{\Omega_0})} e^{-(\frac{\gamma_{2}}{\Omega_0})} d{\gamma_{1}} d{\gamma_{2}}} . \nonumber
\end{align}  
The expression in (\ref{Out_U5}) for  $P \to \infty$, and consequently $\gamma_{\rm th} \to 0$, converges to
  \begin{align} \label{Out_U6}
 {\rm Pr}\big\{q_1^1(t)=q_2^1(t)=0\big\}
  \to \frac{\gamma_{\rm th}^2 }{\Omega_0^2} {\hat \Omega_{IS}} \; \textrm{as} \;P \to \infty,
\end{align} 
where ${\hat \Omega_{IS}}$ is given in (\ref{Otage_S_final}).

 Now, by adding (\ref{Out_U4}), (\ref{Out_U44}), and (\ref{Out_U6}), we obtain the asymptotic  outage probability as in (\ref{eq_out1asl}). This completes the proof. 
\bibliography{litdab}
\bibliographystyle{IEEEtran}

\end{document}